\documentclass[nofootinbib,floatfix,showpacs,preprintnumbers,amsmath,amssymb, prd,11pt]{revtex4}

\usepackage{amsmath}
\usepackage{graphicx}
\usepackage{amssymb}
\usepackage{amsfonts}
\usepackage{latexsym}

\def\beq{\begin{eqnarray}}
\def\eeq{\end{eqnarray}}

\def\diag{\,\mbox{diag}\,}

\renewcommand{\vec}[1]{{\bf #1}}

\renewcommand{\diag}{\,\mbox{diag}\,}

\def\al{\alpha}
\def\be{\beta}

\def\de{\delta}

\def\ka{\kappa}
\def\la{\lambda}
\def\na{\nabla}
\def\pa{\partial}

\def\si{\sigma}
\def\om{\omega}

\def\De{\Delta}
\def\La{\Lambda}

\begin{document}

\preprint{...}

\title{Do we have unitary and (super)renormalizable
Quantum Gravity below Planck scale?}


\author{Filipe de O. Salles}
\email{fsalles@fisica.ufjf.br}
\author{Ilya L. Shapiro}
\email{shapiro@fisica.ufjf.br}
\altaffiliation{Also at Tomsk State
Pedagogical University, Tomsk, Russia.}
\affiliation{Departamento de F\'{\i}sica, ICE,
Universidade Federal de Juiz de Fora, 36036-330, MG, Brazil}

\date{\today}

\begin{abstract}
We explore how the stability of metric perturbations in higher
derivative theories of gravity depends on the energy scale of
initial seeds of such perturbations and on a typical energy
scale of the gravitational vacuum background. It is shown that, at
least in the cases of specific cosmological backgrounds, that the
unphysical massive ghost which is present in the spectrum of such
theories is not growing up as a physical excitation and remains
in the vacuum state, until the initial frequency of the perturbation
is close to the Planck order of magnitude. In this situation the existing
versions of renormalizable and superrenormalizable theories can be
seen as very satisfactory effective theories of Quantum Gravity
below the Planck scale.
\end{abstract}



\maketitle

\section{Introduction}

The situation in Quantum Gravity (QG) has been always shadowed
by the conflict between renormalizability and unitarity. From
one side, General Relativity, which seems to be {\it the theory}
of classical gravity, leads to a non-renormalizable quantum theory
\cite{hove,dene,GorSag}. One can achieve renormalizability by
including general four derivative covariant terms into the action
\cite{Stelle77}, but such terms lead to the unphysical ghost
excitations in the particle spectrum of the theory. Trying to
remove these ghosts from physical spectrum one violates unitarity
of the gravitational $S$-matrix. So, the renormalizable QG is
non-unitary without ghosts,
while the unitary version of QG is non-renormalizable. As a result
of this conflict the idea of Quantum Gravity went far beyond the
conventional approach of perturbatively quantizing the gravitational
field. However, there is an important remaining question: to which
extent one should be afraid of higher derivative ghosts, which are
the source of the difficulty?

The problem of higher derivatives and related instabilities attracted
a lot of attention for a long time. Already in 1850 Ostrogradski
described these exponential type instabilities \cite{Ostrogradski}.
Later on, in 1963, Veltman discussed a process
of quantum scattering of a large-mass negative energy particle and
a much lighter positive-energy particle \cite{Veltman}. In a
simplified qualitative form the net result of this study is that,
typically, the negative-energy particle (massive ghost) gains even
more negative kinetic energy and, consequently, the positive-energy
particle gains more positive kinetic energy. In case of higher
derivative gravity, even if we do not observe ghost due to its
huge mass, there should be intensive gravitons emission, which
can destroy the ``pacific'' classical solution. More recently,
the subject was treated both in the framework of QG
\cite{Tomboulis,salstr, Antomb,Johnston,Hawking,AntDuda}, in
classical gravity
\cite{Stelle78,Whitt,Myung,marot,TorTre,Deru,Mannheim,Tkach}
and for the simplified model theories, mainly based on higher
derivative oscillators \cite{Smilga,Nesternko,Woodard-07,Durer,Kruglov}.

One can note that the mentioned approaches are in fact very
different. The QG-based approaches \cite{Tomboulis,Antomb,AntDuda}
are related to the assumption that the ghost pole gains
gauge-dependent imaginary contribution at quantum level,
leading to the unitary $S$-matrix. Unfortunately, the one-loop
result \cite{julton,frts82,avrbar,Gauss} is not sufficient for
checking whether this desirable quantum effect really takes place
or not\footnote{In our opinion, the situation is qualitatively similar also
for the existing non-perturbative methods (see, e.g., \cite{Percacci}),
but the existence of such methods looks very promising.}.
Another ``quantum'' proposal \cite{Hawking} can be described as
an idea to modify Quantum Field Theory formalism such that ghost
always be treated together with graviton and is not regarded as
an independent particle. For a while, it is not clear how to put
this idea into practise.

The classical approaches \cite{Whitt,Myung,TorTre,Deru} are related
to the exploration of stability for a given (cosmological or
black hole) solution. In the cosmological case it is reduced
to the stability with respect to the perturbations of the
conformal factor of the metric (see also \cite{star,asta})
and also to the stability for the gravitational wave - type
perturbations \cite{star83,HHR,wave,Gasperini,GW-Stab}. It is
remarkable that the perturbations in higher derivative theories
do not show, actually, such a strong instabilities as one
would expect in the theory with unphysical ghosts. It is
important to notice that the mentioned works do not deal just
with the linear perturbations, because the last propagate
on a non-trivial metric background.

The purpose of the present contribution is to consider the
relation between the presence of ghosts and gravitational
instabilities in a spirit of effective Quantum Field Theory.
Our consideration will be simple, purely classical and to
some extent close to the one of \cite{Deru} and \cite{Durer}.
We are going to present some arguments in favor of that the
behavior of the gravitational perturbations is closely
related to the presence of ghosts, but only if the energy
scale is sufficient to generate such a ghost. Our
consideration will be based on the linear perturbations on
a non-trivial gravitational background. Our results will not
be conclusive and should be seen, hopefully, as a contribution
for further investigation of the problem.

The paper is organized as follows. In the next Section we shall
briefly review the reasons to introduce higher derivative terms
and consequently massive unphysical ghosts in the quantum theory.
Sect. 3 includes derivation of the equation for a low-energy
gravitational wave on an arbitrary low-energy gravitational
background and qualitative discussion about the possible effect
of such a background on the time evolution of the gravitational
wave modes. In Sect. 4 the analysis of the metric perturbations
is performed in the relatively simple cases of cosmological
background, for renormalizable and super-renormalizable versions
of the higher derivative theory of gravity, respectively. We show
that the explosive nature of ghosts really takes place, but only
for the initial frequencies of the Planck order of magnitude.
At the same time, nothing like this can be observed for smaller
energies of gravitational perturbations. Finally, in Sect. 5 we
draw our conclusions and discuss possible continuations of this
work.

\section{General situation with massive unphysical ghosts}

One can start by formulating a few general questions concerning
higher derivative ghosts, e.g., as follows:
\ {\it (i)}  Can we survive without them?
\ {\it (ii)}  What is really bad in these ghosts?
\ {\it (iii)}  Can we somehow get rid of them?
\ \
Let us start from the beginning and show that the answer to
the first question {\it (i)} is negative.

\subsection{Can quantum theory survive without gravitational higher derivatives?}

In order to understand why we need higher derivatives in the
gravitational action, one has to start with the relatively
simple situation when only matter fields are quantized and
gravity is a classical background. In this semiclassical
theory one has to introduce the action of vacuum, which is
a functional of the external classical metric. It is well-known
for a long time \cite{UtDW} (for general proofs see, e.g.,
\cite{buch84,Toms83,RenCurved}) that such a theory may be
renormalizable, but only if one introduce the following terms
into the classical action of vacuum:
\beq
S_{vac}\,=\,S_{EH}\,+\,S_{HD}\,,
\label{vac}
\eeq
where
\beq
S_{EH}\,=\,-\,\frac{1}{16\pi G}\,
\int d^4x\sqrt{-g}\,\left\{R + 2\La \,\right\}
\label{EH}
\eeq
is the Einstein-Hilbert term with a cosmological constant and
\beq
S_{HD}=\int d^4x\sqrt{-g}\left\{
a_1C^2 + a_2E + a_3{\Box}R + a_4R^2\right\}
\label{HD}
\eeq
includes higher derivative terms. Here we used the following
notations:
\beq
C^2 &=& R_{\mu\nu\al\be}^2 - 2R_{\al\be}^2 + \frac{1}{3}\,R^2\,,
\nonumber
\\
E &=& R_{\mu\nu\al\be}R^{\mu\nu\al\be} -4 \,R_{\al\be}R^{\al\be} + R^2
\label{CE}
\eeq
for the square of the Weyl tensor and for the Lagrange density of
the Gauss-Bonnet topological term  (Euler density) in \ $d=4$.

The sufficiency of the higher derivative terms (\ref{HD}) for
renormalizability has been consequently proved in a formal way
(see, e.g., \cite{book} for introduction and further references).
The most difficult part is to prove that the diffeomorphism
invariance is preserved at quantum level, and this can be done,
including the case when non-covariant gauges are used for the
background metric \ $g_{\mu\nu}=\eta_{\mu\nu} + h_{\mu\nu}$
\cite{RenCurved}. Furthermore, one has to remember that
all possible UV counterterms are local expressions.
After that the problem reduces to the evaluation of the
superficial degree of divergence in the diagrams with internal
lines of matter fields and external lines of \ $h_{\mu\nu}$.
The theory which is renormalizable in flat space-time has only
mass dimension-four logarithmic divergences.
An important observation is
that adding external lines of \ $h_{\mu\nu}$ does not increase
the degree of divergence (see, e.g., \cite{PoImpo} for more
detailed consideration). Therefore, only dimension-four divergences
will emerge in the same theory, even in curved space. This means
one has to introduce all such terms at the classical level,
that is why we need all terms of (\ref{vac}) in the vacuum
action.

One has to note the great importance of higher derivative
terms (\ref{HD}) for the most important applications of
semiclassical theory. For example, the Hawking radiation
\cite{Hawking-BH} and the general version of Starobinsky
inflation \cite{star} can be derived from the conformal
anomaly \cite{chfu}, and the last results from the
renormalization of the terms (\ref{HD}).

In quantum gravity the higher derivative term with $C^2$ in
(\ref{HD}) means massive ghost, a spin-two particle with
negative kinetic energy. This leads to the problem with
unitarity, at least at the tree level.  But, in the
semiclassical theory gravity is external and the unitarity
of the gravitational \ $S$-matrix may be not considered
really important. The consistency conditions in this case
can include  existence of physically reasonable solutions
and their stability under small metric perturbations.
We shall discuss the relation of such a stability to
the presence of massive ghost in what follows.

Let us now consider the situation in the quantum theory of
gravitational field. Once again, one can prove that the
diffeomorphism invariance is preserved at quantum level (see,
for example, consideration in \cite{Stelle77} which can be
generalized to a wide class of the theories of gravity).
The evaluation of the superficial degree of divergence
\ $D$ \ for a Feynman diagram of the field \ $h_{\mu\nu}$ \
can be performed by means of the general formula
\beq
D + d \,=\,\sum\limits_{l_{int}}(4-r_l)
\,-\,4n \,+\,4\,+\,\sum\limits_{\nu}K_\nu\,,
\label{D+d}
\eeq
with an additional topological relation.
\beq
l_{int} = p + n - 1\,.
\label{top}
\eeq
Here $d$ is the number of derivatives on external lines of
the diagram, \ $r_l$ \ is the power of momenta in the
inverse propagator of internal line, \ $n$ \ is the number
of vertices and \ $K_\nu$ \ is the power of momenta in a
given vertex. In equation (\ref{top}) $l_{int}$ and $p$ are
number of internal lines and loops, correspondingly. $D=0$
case corresponds to the logarithmic divergences and then
$d$ indicates the number of derivatives in the requested
counterterms.

For the quantum version of General Relativity we have
\ $r_l=2$ \ and \ $K_\nu=(2,0)$. It is easy to see from
(\ref{D+d}) and (\ref{top}) that the final expression
for the logarithmic divergences is \ $d = 2+2p$, and
this means the theory is not renormalizable.

If we start from the action (\ref{vac}), which includes
fourth derivative terms  (\ref{HD}), then \ $r_l=4$
\ and \ $K_\nu=(4,2,0)$. It is easy to see that the
maximal power of derivatives in the logarithmic counterterms
is $d=4$, so this theory is renormalizable at all loop orders.

One can introduce more derivatives, by considering the
action \cite{highderi}
\beq
S &=& S_{EH} \,+\,\int d^4x\sqrt{-g}\,\Big\{
a_1R_{\mu\nu\al\be}^2 + a_2R_{\mu\nu}^2 + a_3R^2 + \, ...
\nonumber
\\
&+& b_1 R_{\mu\nu\al\be} \Box R^{\mu\nu\al\be}
+ b_2 R_{\mu\nu}\Box R^{\mu\nu}
+ b_3 R\Box R
+ \, {\cal O}(R_{...}^3) \,+
\nonumber
\\
&+& c_1 R_{\mu\nu\al\be} \Box^k R^{\mu\nu\al\be}
+ c_2 R_{\mu\nu}\Box^k R^{\mu\nu}
+ c_3 R\Box^k R
\,+\, ...\,+ \, {\cal O}(R_{...}^{k+2})
\Big\}\,.
\label{higher}
\eeq
For this theory, in general case, we have $r_l=4+2k$ \ and \
$K_\nu=(0,2,...,4+2k)$. By means of Eqs.  (\ref{D+d}) and
(\ref{top}), for the logarithmic divergences one has
\ $d\,=\,4\,+\,k(1-p)$. This formula has three important
consequences. First, the theory is superrenormalizable
for $k\geq 1$ and only one-loop divergences are present
for $k\geq 3$. Second, all divergences are fourth-derivative
ones or less. This means that most of the terms in
(\ref{higher}) are not renormalized. Third, the zero-derivative,
second-derivative and four-derivative counterterms depend
on the choice of coefficients in the highest derivative terms.
Let us note, incidentally, that the power-counting in the
popular Ho\v{r}ava-Lifshits gravity \cite{Horava} is exactly
the same as the one described above. This means that the
four-derivative in time logarithmic divergences in this
theory are very likely to show up, but maybe can be cancelled
by a special fine-tuning of highest derivative terms. Only
explicit calculation can demonstrate whether this really
happens or not, and one-loop calculation would be sufficient
for $k\geq 3$. Anyway, as far as a target is {\it pure} QG,
the Ho\v{r}ava-Lifshits gravity has a good chance.
At the same time, there is a more serious difficulty
related to the contribution of matter fields. If the
Lorentz violation in the matter sector is not assumed,
these fields will always produce covariant $R^2_{\mu\nu}$-type
divergences at all loop orders and hence it is unclear how one
can construct a theory without fourth order time derivatives.
Some support for this consideration comes, also, from the direct
calculations for a scalar field in \cite{DiMaz}.

The massive ghosts are still present in the theory
(\ref{higher}). For the case of real poles we have
\cite{highderi}
\beq
G_2(k) = \frac{A_{0}}{k^2} +
 \frac{A_{1}}{k^2 + m_1^2} + \frac{A_{2}}{k^2 + m_2^2}
+ \cdots + \frac{A_{N+1}}{k^2 + m_{N+1}^2}\,,
\label{poles}
\eeq
where the signs alternate
\beq
A_j\cdot A_{j+1} < 0
\label{alter-ego}
\eeq
for any sequence with growing real masses
\beq
0 < m_1^2 < m_2^2 < m_3^2 < \cdots < m_{N+1}^2\,.
\label{masses}
\eeq
In principle, it would be interesting to explore the cases
of imaginary and negative poles (e.g., looking for some kind
of a see-saw mechanism for the ghost poles), but we shall
leave such a consideration for the future work. In the present
paper our attention will be restricted by the case (\ref{poles})
and discuss the relation between the presence of ghosts and
gravitational instabilities of the vacuum state of the theory.

Looking at the expression (\ref{poles}) one can see that
this theory has one (in the case $k=1$) or more (for $k\geq 2$)
ghost degrees of freedom in the tensor sector. We conclude
that, in general, the price of (super)renormalizability is the
presence of ghosts (see also additional discussion of this
issue and further references in \cite{torsi}). However, from
the general perspective, the most important argument in favor
of higher derivatives comes from the quantization of matter
fields. Taking into account the importance of $S_{HD}$ in
(\ref{HD}) for constructing renormalizable action of vacuum
of quantum matter fields, it is really difficult to see how
one can achieve a consistent theory without covariant higher
derivatives, so it is worthwhile to take the presence of ghosts
seriously and see how we can deal with them.

\subsection{Can one get rid of massive ghosts?}

Massive ghosts are tensor (spin-two) massive states with
negative kinetic energy. The corresponding components of
the propagator do not depend on the gauge fixing and can
be seen as physical degrees of freedom. Creation of the
particle with negative kinetic energy from the vacuum state
is not protected by energy conservation, this means that in
the theory with ghosts one should expect continuous creation
of ghosts and a lot of high-energy gravitons (remember that
``our'' ghost has Planck-order mass). Even if we do not see
ghost itself, we are going to observe a huge destructing
outflux of gravitational energy, which is supposed to
explode any classical gravitational solution (see,
e.g., \cite{Woodard-07} for a recent review).

There were, as we have already mentioned in the Introduction,
several interesting attempts to get rid of the massive ghosts.
The most obvious idea is
to assume that the initial \ $|in\rangle$ \ state in the classical
scattering of gravitational perturbations does not contain ghost.
The problem is that, due to the non-polynomial nature of gravity,
the ghost has infinitely complicated interaction to gravitons and,
as a result of this interaction, there should be ghosts in the
\ $|out\rangle$ \ state. Then the theory will be non-unitary.
Let us mention a recent work \cite{Nara}, where it was shown that
the theory {\it without} ghost is unitary. In our opinion this is
not a real solution, because the problem is exactly of how one
can remove the ghost from the spectrum.

One of the interesting ideas is related to the possible
role of quantum corrections on the unphysical massive pole
\cite{Tomboulis,salstr,Antomb}. As we have already discussed
in the introduction, the existing methods do not enable one to
perform non-perturbative analysis which is needed to make final
conclusion about this possibility \cite{Johnston}. Let us note
that the situation can be somehow better in the  superrenormalizable
version of the theory (\ref{higher}), where it is technically
possible to calculate exact $\be$-functions and thus arrive at
the leading approximation to the full quantum-corrected propagator.
We shall leave this possibility for the future work and will
concentrate, instead, on a much simpler, direct approach to
the problem of ghosts and instabilities.

An interesting possibility has been suggested in \cite{Tomb97}
and developed further recently in \cite{Modesto,Koivisto} (see
further references on classical applications therein). The idea
is to continue an expansion in (\ref{higher}) to the infinite order
in derivatives. The expectation is that one can achieve the following
form of the bilinear expansion of the classical action (for simplicity,
we take a flat background here, and assume that an appropriate gauge
fixing term is included):
\beq
S^{(2)} &=& \frac12\,h^{\al\be}\, \big\{
cM_P^2\,\Box \,+\, f(\Box)\big\}\,h_{\al\be}\,,
\label{infi}
\eeq
such that \ $c=const$ \ and \ $f(\Box)$ \ is chosen such that the
sum \ $cM_P^2\,\Box \,+\, f(\Box)$ \ is a specially designed entire
function of the argument \ $\Box$. It is assumed that the resulting
theory is (super)renormalizable and that the propagator of the
gravitational perturbation \ $h^{\al\be}$ \ has a unique pole at
\ $k^2=0$. The idea looks very nice and beautiful, but there are
certain doubts about whether this scheme will work for QG.
\ First, in order to
claim that the theory is (super)renormalizable, one has to arrive
at the Feynman rules for \ $h^{\al\be}$, and for this end one need
to perform quantization of the theory. It is not clear how this can
be done in the non-polynomial in derivatives theory like (\ref{infi}).
Second, in the theory (\ref{infi}) one has both $r_l$ and $K_\nu$
infinite. Therefore, the evaluation of the superficial degree of
divergence (\ref{D+d}) in this theory will produce an indefinite
$(\infty-\infty)$ - type result, so it is unclear what one can say
about this theory being superrenormalizable, renormalizable or
non-renormalizable. In the present case the possibility of
non-renormalizable theory means, in particular,
that the form of the function  \ $f(\Box)$ \ may eventually change
under quantum corrections, such that the massive pole will come
back to the theory. Starting from the expression of the actions like
\beq
S_{iHD} &=& \int d^4x\sqrt{-g}\,\big\{cR
\,+\, R^{\mu\nu} \,h(\Box)\,R_{\mu\nu}
\,+\, R \,h_1(\Box)\,R
\big\}\,.
\label{infi-co}
\eeq
we arrive at the inverse propagator (in momentum space, for the
spin-two sector) of the form
\beq
G^{-1}(k) &=& c_1 k^2 +  k^4 \psi(k^2)\,,
\label{infi-prop}
\eeq
where $\psi(k^2)$ is an analytic function and $c_1\neq 0$. One can
provide an absence of extra poles with \ $k\neq 0$ (real or
complex) in such case, by setting, e.g.,
\beq
c_1 +  k^2 \psi(k^2) \,=\, c_1 \,e^{-\,k^2/M^2}
\label{infi-exp}
\eeq
or in some other similar way \cite{Tomb97}, but it is not obvious
that this form of the function will hold after quantum corrections
are taken into account. Finally, the proposal of  \cite{Tomb97} is
very interesting, but the statement that the theory based on
(\ref{infi}) really solves the conflict  between renormalizability
and unitarity looks a little bit premature and is not clarified,
until now.

A qualitatively distinct approach has been suggested in \cite{Hawking},
it is based on the observation that ghost is not an independent particle,
but rather a companion of the graviton in the linearized gravity. The
separation of different degrees of freedom in higher derivative theories
is a non-trivial issue even in case of linearized theories (see, e.g.,
\cite{Julve-HD}). Needless to say that the situation should be more
complicated in gravity, which has a non-polynomial
interaction structure. However, up to now it is not clear how one
can put the proposal of Ref. \cite{Hawking} into practise, and how
the new quantum theory of gravity should look. Anyway, these two
proposals show that the situation with ghost is not completely
hopeless and should be explored in more details.

Finally, let us mention the literature of avoiding the ghosts in
the models of massive gravity \cite{magra-refs}. An alternative
approach here is to admit that the unphysical ghost may exist,
but it is harmless, because its interaction to the rest of the
particles is non-local and is suppressed by some large parameter
\cite{Maggiore-13}. It looks tentative to find some mechanism
with similar final output for the much more relevant (at least,
in our opinion) case of higher derivatives. In what follows we
consider the possibility that the corresponding ghost exists only
as a vacuum excitation, but never shows up as a physical particle
and, therefore, maybe harmless at the energies below the Planck scale.

\section{Gravitational waves on an arbitrary background}

Let us remember the assumptions which were done to deal with
the ghost problem in higher derivative theory.

{\Large $\bullet$} \
One can draw conclusions about the gravity theory by using
linearized approximation. The $S$-matrix of gravitons should
be the main object of our interest.

{\Large $\bullet$} \  Ostrogradsky instabilities
\cite{Ostrogradski} or Veltman scattering \cite{Veltman} are
relevant independent on the energy scale, in all cases they
produce run-away solutions and the Universe explodes.

There is a simple way to directly check most of these assumptions
at once. Let us take a higher derivative theory of gravity and
verify the stability with respect to the linear perturbations
on some, physically interesting, classical solution. If the
mentioned assumptions are correct, we will observe rapidly
growing modes even for the low-energy (i.e., low-curvature)
background. However, if there are no growing modes at the linear
level, there will not be such modes even at higher orders.
Let us remember that the ghost problem is a tree-level one,
and therefore we do not need to worry about loop effects.
Moreover, according to the known mathematical theorem
\cite{Math}, if the system is stable with respect to linear
fluctuations, it will be stable at the non-linear level too,
at least for the sufficiently small amplitudes of
perturbations.

Finally, our general purpose is to explore the time dynamics
of the gravitational waves on an arbitrary ``low-energy''
background, in a higher derivative theory of gravity. In
what follows we shall start from the theory (\ref{vac})
on a general background and show that there are some
arguments in favor of its irrelevance for the sufficiently
low energy fluctuations. In the consequent section we shall
deal with the reduced problem and identify the relations
between the presence of growing modes and existence of
massive ghosts on the cosmological background.

\subsection{Riemann Normal Coordinate expansions}

Let us consider the fourth-derivative theory (\ref{vac}) and set
to zero the cosmological constant. This is justified when we are
interested in the behaviour of the gravitational waves, because
the cosmological constant is irrelevant at the distances much
smaller than the size of the Universe. The action which we
will deal with can be cast into the form
\beq
S_{4dQG}\,=\,
\int d^4x\sqrt{-g}\,\left\{
-\,\frac{M_P^2}{16\pi}\,R\,
\,+\,a_1C^2 + a_2E + a_3{\Box}R + a_4R^2\right\}\,.
\label{4dQG}
\eeq
The unique dimensional parameter in this theory is Planck mass
\ $M_P$, because all other coefficients are dimensionless. Of
course, \ $a_k$ \ are arbitrary parameters and we can choose
them to be as great as we like, but let us make a moderate
choice, assuming that the values of  \ $a_k$ \ are close, in
the orders of magnitude, to unity. Then the  Planck mass \ $M_P$ \
defines the unique scale of the theory. This means that all
those quantities which are much smaller than \ $M_P$ \ are very
small in this theory. One can note that this feature has been
extensively used in establishing the effective approach to QG
\cite{don94}.

The low-energy approach to the dynamics of gravitational
perturbations on an arbitrary metric background means that
the following inequalities are satisfied:
\beq
\left| R_{\mu\nu\al\be} \right| \ll M_P^2
\qquad \mbox{and} \qquad
{\vec k}^2 \ll M_P^2\,,
\label{ineq}
\eeq
where $R_{\mu\nu\al\be}$ are components of the Riemann
tensor of a background and ${\vec k}$ is a wave vector
for the perturbation.

The equation of our interest is
\beq
&& H^{\mu\nu , \,\al\be}\,{\bar h}^{\bot}_{\al\be} \,=\, 0\,,
\label{eq1}
\\
\mbox{where} && \,\,
H^{\mu\nu , \,\al\be} \,=\,
\frac{\de^2\,S_{4dQG}}{\de g_{\mu\nu}\,\de g_{\al\be}}\,.
\nonumber
\eeq
The gauge fixing term is irrelevant, since we are interested
only in the traceless and completely transverse components of
the gravitational perturbation ${\bar h}^{\bot}_{\al\be}(x)$,
which will be denoted $h_{\al\be}$ in what follows. We will
assume that $h_{\al\be}$ satisfies the constraints
\beq
h_{\al\be}\,g^{\al\be}=0
\quad
\mbox{and}
\quad
\na^\al\,h_{\al\be}=0\,.
\label{const}
\eeq

As an illustration, let us write separately
the zero-order in curvature terms in (\ref{eq1}) as
\beq
a_1\,\Big(
\Box^2 \,-\,\frac{M_P^2}{32\pi\,a_1}\,\Box\Big)
\,h^{\al\be} \,=\,0\,,
\label{eq1-pure}
\eeq
which corresponds to the mass of the ghost
\ $m_2 = M_P/\sqrt{32\pi a_1}$.

The full equation includes the
terms (\ref{eq1-pure}) and also terms linear and quadratic in
curvature. One can easily obtain this equation from the works
on HDQG, e.g., \cite{book} or \cite{Gauss}. In the first order in
curvature and taking into account (\ref{const}), this equation has
the form (\ref{eq1}) with \footnote{In these expressions, the
symmetrization over the pairs of indices $(\mu\nu)$ and $(\al\be)$
is assumed. The complete forms including second order terms can
be found in Ref. \cite{Gauss}.}
\beq
H_{\mu\nu , \,\al\be} \,=\,
-\,\frac{a_1}{2}\,\de_{\mu\nu , \,\al\be}\Box^2
+ D^{\rho\la}_{\quad \mu\nu , \al\be}\na_\rho\na_\la
\,+\,W_{\mu\nu , \,\al\be},,
\label{fulleq}
\eeq
where
\beq
D^{\rho\la}_{\quad \mu\nu , \al\be}
&=&
2a_1 g_{\nu\be} R_{\al\,\cdot\,\cdot\,\mu}^{\,\,\,\rho\la}
+ a_1 g^{\rho\la} \big( 2g_{\nu\be}  R_{\al\mu}
- R_{\mu\al\nu\be}\big)
+ \Big(\frac{M_P^2}{64\pi}-\frac{a_1}{6} R
-\frac{a_4}{2} R\Big) g^{\rho\la}
 \de_{\mu\nu , \al\be}\,;
\nonumber
\\
W_{\mu\nu , \,\al\be}
&=&
\frac{M_P^2}{64\pi}\,
\big(R_{\mu\al\nu\be}
+ 3 \,R_{\mu\al}\,g_{\nu\be}
- R\,\de_{\mu\nu ,\al\be}\big)\,.
\label{feqdets}
\eeq

The reason to keep only linear terms  in curvature is due to
our interest in the behavior of metric perturbations in
equation (\ref{eq1}) when both background and perturbations
have typical energies much smaller that the Planck scale.
This means, in particular, that we can ignore all
${\cal O}(R^2_{....})$-terms. Of course, it would
be interesting to explore higher orders, at some point,
but in the present work we will try to make calculations as
simple as possible.

It is natural to use some technique which enables one
to treat curvature tensor components as small perturbations.
The covariant formalism of this kind is based on the Riemann
normal coordinates \cite{Petrov}. This approach is traditionally
used for describing the propagator \cite{BunPar}, in our case
for gravitons.
The method is also useful in other situations, mainly related to
the evaluation of loop effects \cite{parker-toms,CorPot}, but
now we intend to discuss only the tree-level approximation.

The normal coordinates method assumes an expansion around a
chosen point in the space-time, let's call it $P(x^{\prime \mu})$.
The quantities corresponding to this point will be labeled by small
zero, for instance the metric is
\ ${\stackrel{\mbox{\tiny o}}{g}}_{\al\be}$. Also,
we shall need curvature tensor and its covariant derivatives at
this point. The nice feature of normal coordinates is that the
coordinate lines are specially designed geodesic lines and
an expansion can be done covariantly with respect to the
point $P$. The deviation from the point $P$ is parameterized
by the quantities $y^\mu=x^\mu-x^{\prime \mu}$, which are zero
at $P$. As far as we consider the components of the curvatures
to be small, the consideration can be restricted by the
first order terms. For the sake of generality we shall
perform also part of the expansion until second order,
the corresponding results are settled in Appendix A.

The expansion for the metric has the form
\beq
g_{\al\be}(y)
&=& {\stackrel{\mbox{\tiny o}}{g}}_{\al\be}
\,-\,\frac13\,{\stackrel{\mbox{\tiny o}}{R}}_{\mu\al\nu\be}
\,y^\mu\,y^\nu
+\,...\,.
\label{metric2}
\eeq
One can always choose the metric in the expansion point to be
Minkowski one, \ ${\stackrel{\mbox{\tiny o}}{g}}_{\al\be} =
\eta_{\al\be}$. For the Christoffel symbol one has
\beq
\nonumber
\Gamma^{\lambda}_{\alpha\beta}
&=&
\frac{2}{3}\,
{\stackrel{\mbox{\tiny o}}{R}}\,^{\la}_{\,\cdot\,(\al\be)\nu} \,y^{\nu}
\, + \,...\,.
\label{Gamma}
\eeq

Let us start from the normal coordinates expansion for
\ $\Box\,h^{\al\be}$. The expansion represents a power
series in both curvature components
${\stackrel{\mbox{\tiny o}}{R}}\,^{\la}_{\,\cdot\,\al\be\nu}$
and $\,y^{\mu}$. In what follows we label by $A^{(n)}$ the
order $n$ of the expansion in \ $y^\mu$ for the quantity $A$,
for instance
\beq
\Box\,h^{\alpha\beta}\,=\,(\Box\,h^{\alpha\beta})^{(0)} +
(\Box\,h^{\alpha\beta})^{(1)} + (\Box\,h^{\alpha\beta})^{(2)}\,+\,\,..\,,
\label{uravn}
\eeq
where the dots indicate to the omitted terms of higher orders in
$y^\mu$ and of higher orders in curvature tensor and its covariant
derivatives at the point $P$.
Direct calculation yields the following results up to the second
order in $y^\mu$:
\beq
(\Box\,h^{\alpha\beta})^{(0)}
&=&
\eta^{\mu\nu}\Big[\partial_\mu \partial_\nu\, h^{\alpha\beta}
- \frac13\,{\stackrel{\mbox{\tiny o}}{R}}\,^{\al}_{\,\cdot\,\nu\la\mu}
\,h^{\lambda\beta}
- \frac13\,{\stackrel{\mbox{\tiny o}}{R}}
\,^{\be}_{\,\cdot\,\nu\la\mu}\,h^{\al\la}\Big]\,,
\label{h0}
\\
(\Box\,h^{\alpha\beta})^{(1)}
&=&
-\frac{4}{3}\,\eta^{\mu\nu}
\Big[{\stackrel{\mbox{\tiny o}}{R}}
\,^{\al}_{\,\cdot\,(\nu\la)\tau}\,\pa_\mu\,h^{\la\beta}
\,+\, {\stackrel{\mbox{\tiny o}}{R}}\,^{\be}_{\,\cdot\,(\nu\la)\tau}
\,\pa_\mu\,h^{\al\la}\Big]\,y^{\tau}\,,
\label{h1}
\\
(\Box\,h^{\al\be})^{(2)}
&=&
\frac{1}{3}\,
{\stackrel{\mbox{\tiny o}}{R}}
\,^{\mu\,\,\,\nu}_{\,\cdot\,\tau\,\cdot\,\rho}
\,\big(\pa_\mu \pa_\nu h^{\al\be}\big)\,y^\tau\,y^\rho\,.
\label{h2}
\eeq


\subsection{Zero-order approximation}

The next step would be to make a Fourier transformation in the
spatial sector,
\beq
h_{\mu\nu}({\vec r},t)
&=&
\int \,\frac{d^3k}{(2\pi)^3}\,
h_{\mu\nu}({\vec k},t) \,e^{i{\vec k}\cdot{\vec r}} \,.
\label{Fu}
\eeq
As a useful approximation, we can treat the wave vector ${\vec k}$
as constant and will be therefore interested only in the time
evolution of the perturbation \ $h_{\mu\nu}$. The validity of
such a treatment is restricted to the long-wave perturbations,
where we assume that the modes \ $h_{\mu\nu}({\vec k},t)$ \ have
independent time dynamics. This treatment enables one to trade
the complicated partial differential equation (\ref{eq1}) to
the much simpler ordinary differential equations for individual
modes. Since in the theory under discussion the unique scale
parameter is given by the Planck mass, the long wavelength is
just the one which is larger than the Planck length.

Let us now see what the approximation of independent modes
\ $h_{\mu\nu}({\vec k},t)$ \ means, from the practical side.
Looking at the Eqs. (\ref{h0}), (\ref{h1}) and (\ref{h2}), it
is clear that the equation (\ref{eq1}) has two complications:
those related to the derivatives like \
$\pa h^{\al\be}({\vec r},t)/\pa y^\mu$, and also
related to the factors of $y^\mu$. Obviously,
$\pa h^{\al\be}({\vec r},t)/\pa y^\mu$ reduce, after using
(\ref{Fu}), to the $ik_\mu h^{\al\be}({\vec k},t)$. The
treatment of the factors of $y^\mu$ is a bit more complicated
and goes as follows:
\beq
\int \,\frac{d^3k}{(2\pi)^3}\,y^\mu\,
h^{\al\be}({\vec k},t) \,e^{i{\vec k}\cdot{\vec r}}
&=&
\int \,\frac{d^3k}{(2\pi)^3}\,
h^{\al\be}({\vec k},t) \,\frac{\pa}{i \pa k_\mu}\,
e^{i{\vec k}\cdot{\vec r}}
\,.
\label{Fu-1}
\eeq
One can integrate by parts in the last expression. The
surface term at infinity can be neglected, because we can
assume $h^{\al\be}(\left|{\vec k}\right| \to \infty) \to 0$,
since all perturbations are suppressed beyond Planck scale.
In this way we arrive at the relation
\beq
\int \,\frac{d^3k}{(2\pi)^3}\,y^\mu\,
h^{\al\be}({\vec k},t) \,e^{i{\vec k}\cdot{\vec r}}
&=&
\int \,\frac{d^3k}{(2\pi)^3}\,
e^{i{\vec k}\cdot{\vec r}}
\,\frac{\pa}{i \pa k_\mu}\,h^{\al\be}({\vec k},t)\,.
\label{Fu-2}
\eeq
At that point we conclude that the expansion in normal
coordinates $y^\mu$ means an expansion of modes
$h^{\al\be}({\vec k},t)$ in the series in $k^\mu$. In the
simplest possible approximation we assume that the modes
do not depend on $k^\mu$, that is
$h^{\al\be}({\vec k},t)=h^{\al\be}(t)$. This means we can
restrict the consideration to the zero-order approximation
in $y^\mu$ in the equation (\ref{eq1}).

For the $\Box^2$ term one can write
\beq
(\Box^2\,h^{\alpha\beta})^{(0)}\,=\,
\eta^{\mu\nu}\,\Big\{\Big[\partial_{\mu}\partial_{\nu}\,
(\Box\,h^{\alpha\beta})\Big]^{(0)} - \frac{1}{3}
{\stackrel{\mbox{\tiny o}}{R}}\,^{\alpha}_{\,\,\,\nu\tau\mu}
(\Box\,h^{\tau\beta})^{(0)} - \frac{1}{3}
{\stackrel{\mbox{\tiny o}}{R}}\,^{\beta}_{\,\,\,\nu\tau\mu}
(\Box\,h^{\alpha\tau})^{(0)}\Big\}\,.
\label{zero}
\eeq
Let us introduce the following notations for the
expansions (\ref{h1}) and  (\ref{h2})\footnote{Higher order expressions for
$\De_{\,\,\,\,\,\chi}^{\alpha\beta}$ and $\La_{\,\,\,\,\,\chi\om}^{\al\be}$
can be found in the Eqs. (\ref{h1-A}) and (\ref{h2-A}) in Appendix A.}:
\beq
(\Box\,h^{\alpha\beta})^{(1)}
&=&
\De_{\,\,\,\,\,\chi}^{\alpha\beta}\,
\,y^{\chi}\,,
\nonumber
\\
(\Box\,h^{\alpha\beta})^{(2)}
&=&
\La_{\,\,\,\,\,\chi\om}^{\al\be}
\,y^{\chi}\,y^{\om}\,.
\eeq
After a very small algebra we obtain
\beq
(\Box^2\,h^{\alpha\beta})^{(0)}
&=&
\eta^{\mu\nu}\pa_\mu\pa_\nu\,(\Box\,h^{\alpha\beta})^{(0)}
\,+\,
\eta^{\mu\nu}\Big[
2\pa_\nu\,\De_{\,\,\,\,\,\mu}^{\al\be}
+ 2\La_{\,\,\,\,\,\nu\mu}^{\al\be}
\nonumber
\\
&-&\frac{1}{3}{\stackrel{\mbox{\tiny
o}}{R}}\,^{\alpha}_{\,\,\,\nu\tau\mu}
(\Box\,h^{\tau\beta})^{(0)} - \frac{1}{3}{\stackrel{\mbox{\tiny
o}}{R}}\,^{\beta}_{\,\,\,\nu\tau\mu}
(\Box\,h^{\alpha\tau})^{(0)}\Big]\,,
\label{h4}
\eeq
where $(\Box\,h^{\alpha\beta})^{(0)}$ has been defined in
(\ref{h0}). Taking this
together with (\ref{feqdets}),  we arrive at the expression
\beq
H_{\mu\nu , \,\al\be}
&=&
-\,\frac{a_1}{2}\,\de_{\mu\nu , \,\al\be}(\Box^2\,h^{\alpha\beta})^{(0)}
\,+\,
2a_1\eta_{\nu\be}\,
{\stackrel{\mbox{\tiny o}}{R}}
\,_{\al\,\cdot\,\cdot\,\mu}^{\,\,\,\,\rho\la}
\,\pa_\rho\pa_\la
\label{URA-H}
\\
&+&
\Big[
\Big(
\frac{a_1}{6}\,{\stackrel{\mbox{\tiny o}}{R}}
+ \frac{a_4}{2}\,{\stackrel{\mbox{\tiny o}}{R}}
+ \frac{M_P^2}{64\pi} \Big)\,\de_{\mu\nu ,\,\al\be}
\,+\, 2a_1\,\eta_{\nu\be}\,{\stackrel{\mbox{\tiny o}}{R}}_{\al\mu}
- a_1{\stackrel{\mbox{\tiny o}}{R}}_{\mu\al\nu\be}
\Big]\Box
\nonumber
\\
&+&
\frac{M_P^2}{64\pi}\,\big(
{\stackrel{\mbox{\tiny o}}{R}}_{\mu\al\nu\be} + 3 \,\eta_{\mu\al}
\,{\stackrel{\mbox{\tiny o}}{R}}_{\nu\be}
- {\stackrel{\mbox{\tiny o}}{R}}\,\de_{\mu\nu ,\,\al\be}\big)\,.
\nonumber
\eeq
By replacing (\ref{h4}) into the last formula, we obtain the
equation for the metric perturbation in the zero-order
approximation in $y^\mu$,
\beq
\Box^2h_{\mu\nu}
&-&
\frac13\,\big({\stackrel{\mbox{\tiny o}}{R}}_{\la\mu}\Box h_\nu^\la
+ {\stackrel{\mbox{\tiny o}}{R}}_{\la\nu}\Box h_\mu^\la\big)
\,+\,
\frac43\,\big(
{\stackrel{\mbox{\tiny o}}{R}}
\,^{\la\rho\tau}_{\,\cdot\,\cdot\,\cdot\,\mu}\,\pa_\rho\pa_\tau\,
h_{\nu\la}
+ {\stackrel{\mbox{\tiny o}}{R}}\,^{\la\rho\tau}_{\,\cdot\,\cdot\,\cdot\,\nu}
\,\pa_\rho\pa_\tau\, h_{\mu\la}
\big)
\nonumber
\\
&-&
2\,{\stackrel{\mbox{\tiny o}}{R}}
\,_{\la\,\cdot\,\cdot\,\mu}^{\,\,\,\,\rho\tau}
\pa_\rho\pa_\tau h^\la_{\nu}
\,-\,2\,{\stackrel{\mbox{\tiny o}}{R}}
\,_{\la\,\cdot\,\cdot\,\nu}^{\,\,\,\,\rho\tau}
\pa_\rho\pa_\tau h^\la_{\mu}
\,-\, 2\,{\stackrel{\mbox{\tiny o}}{R}}
\,_{\tau\mu}\Box\,h^\tau_{\nu}
\,-\, 2\,{\stackrel{\mbox{\tiny o}}{R}}
\,_{\tau\nu}\Box\,h^\tau_{\mu}
\nonumber
\\
&+&
2\,{\stackrel{\mbox{\tiny o}}{R}}
\,_{\mu\rho\nu\tau}\Box\,h^{\rho\tau}
\,+\, \frac23\, {\stackrel{\mbox{\tiny o}}{R}}^{\rho\tau}\,
\pa_\rho\pa_\tau h_{\mu\la}
\,+\,\frac{a_1+3a_4}{a_1}\,{\stackrel{\mbox{\tiny o}}{R}}\,\Box  h_{\mu\nu}
\nonumber
\\
&-&
\frac{M_P^2}{32\pi\,a_1}\,\Big[
\big(\Box - {\stackrel{\mbox{\tiny o}}{R}}\big) h_{\mu\nu}
+ \big({\stackrel{\mbox{\tiny o}}{R}}_{\mu\la\nu\tau} + 3 \,\eta_{\mu\la}
\,{\stackrel{\mbox{\tiny o}}{R}}_{\nu\tau}\big)h^{\la\tau}
\Big]\,=\,0\,.
\label{URA}
\eeq
We note that (\ref{URA}) is a flat-space differential equation,
which depends on the curvature tensor components in a given
point $P$, \ $\stackrel{\mbox{\tiny o}}{R}^\al_{\,\cdot\,\be\mu\nu}$.
In particular, here we assume flat d'Alembertian operator,
\ $\Box=\eta^{\rho\tau}\pa_\rho\pa_\tau$. Of course, the complete
expression is an infinite series expansion in both $k^\mu$ and
${\stackrel{\mbox{\tiny o}}{R}}_{\mu\nu\al\be}$, so equation (\ref{URA})
is just the lowest-order nontrivial approximation to it. The equation
(\ref{URA}) is a generalization of the basic equation (\ref{eq1-pure})
and the difference between the two is represented by the terms
linear in curvature which are partially hidden and partially
omitted in (\ref{eq1-pure}). The investigation of
the time dynamics of $h_{\mu\nu}$ with a constant ${\vec k}$
can be performed on the basis Eq. (\ref{URA}). One can expect
that the non-linearities, presented by  non-trivial background
will be responsible for a relatively small corrections to the
dynamics of (\ref{eq1-pure}) in flat space. This statement can
be correct or not and at the moment we are unable to give a
definite answer on the basis of equation (\ref{URA}). Instead we
shall perform partial verification of this statement for the case
of cosmological background, in the next section.

Equation (\ref{URA}) contains relevant information about the evolution
of traceless and transverse mode of the metric perturbation in the
regime
$\,\big|{\stackrel{\mbox{\tiny o}}{R}}_{\al\be\tau\la} \big|
\ll M_P^2$.
We postpone the analysis of this complicated equation for the future
work. In the next section, we shall consider another approximation,
which is not related to the expansion around the flat space. To some
extent, the results of this consideration will justify the system of
approximations which were used in deriving equation (\ref{URA}).

\section{Perturbations on the cosmological background}

Let us now turn to a very different approach and consider
cosmological background metric. In this case the
consideration is not related to the weak-curvature approximation,
but the background is of course a very special on. Anyway,
this consideration can be useful in collecting evidence in
favor of (in)stability of the background in higher derivative
gravity theory. One has to note that classical cosmological
solutions can be very different and hence the problem is
technically not completely trivial.

The consideration of metric perturbations
in higher derivative theories on a cosmological backgrounds has
been previously studied in Refs. \cite{star83,wave,HHR} for the
particular case of inflationary (dS) background and recently
in \cite{GW-Stab} for more general FLRW metrics, namely radiation
and dust-dominated cases. In all these works the equations were
derived on the basis of higher derivative theory with semiclassical
corrections, and in all cases no instabilities were detected.
Here we restrict our attention to the purely classical theory
(\ref{4dQG}). Compared to the previous publications we shall
extend the set of initial conditions and finally discover the
unstable case, exactly in the situation which will confirm the
main assumptions formulated in the previous sections.

We consider perturbations
\beq
g_{\mu\nu}\,=\,g_{\mu\nu}^{0} + h_{\mu\nu}
\label{perturba}
\eeq
over an isotropic and homogeneous
cosmological background,
\beq
g^{0}_{\mu\nu}\,=\,\diag\big\{1,\,-\delta_{ij}\,a^2(t)\}\,.
\label{back}
\eeq
One pertinent observation is in order here. The action (\ref{4dQG})
without cosmological constant has only one term which can affect the
solution for the scale function $a(t)$ of the cosmological background.
Remember that the Weyl tensor is zero for the metric (\ref{back}) and
the Gauss-Bonnet term does not contribute to the equations of motion
in $d=4$. By the end of the day the only relevant higher derivative
term for the background is\footnote{It is interesting that the
$a_1 C^2$-term is much more relevant for the metric perturbations
than the $a_4 R^2$-term, so the situations for the background and
for metric perturbations are just opposite.} $a_4 R^2$.
Then, as far as we consider
the low-energy situation with $\left|R\right| \ll M_P^2$, the classical
solutions of GR can be seen as a precise approximations for
the theory (\ref{4dQG}). For this reason we shall consider the
metric perturbations over the background (\ref{back}) with $a(t)$
corresponding to the standard cosmological solutions of GR, such
as matter-dominated, radiation-dominated Universe and to the
exponential case. In the last case the accelerated expansion
is due to the cosmological constant only.

The initial conditions for the perturbations will be chosen
to originate from the fluctuations of free quantum fields. The
spectrum is identical to a scalar quantum field in Minkowski
space (see, e.g., \cite{birdav}),
\beq
h(x,\eta) = h(\eta)\,e^{\pm i\vec k.\vec r} \quad ,
\quad h(\eta) \propto \frac{e^{\pm ik\eta}}{\sqrt{2k}} \,.
\eeq
where we employed the conformal time $\eta$, \ $a(\eta)d\eta = dt$,
\ $\vec{k}$ is the wavenumber vector and \ $k=|\vec{k}|$.
A normalization constant is not necessary for the case of linear
perturbations. Initial amplitudes are supposed to have quantum
origin and depend on $\vec k$ according to
\beq
h_0 \propto \frac{1}{\sqrt{2k}} \quad , \quad \dot h_0 \propto
\sqrt{\frac{k}{2}} \quad,
\quad \ddot h_0 \propto \frac{k^{3/2}}{\sqrt{2}} \quad , \quad
{\stackrel{...} h}_0
\propto
\frac{k^{5/2}}{\sqrt{2}} \,,
\label{initi}
\eeq
where the derivatives are taken with respect to the cosmic time.
Let us stress that the vacuum stability is related to the asymptotic
behavior of perturbations at $t \to \infty$ and therefore the
choice of initial conditions is, to a great extent, irrelevant.
However, all plots presented below correspond to Eqs. (\ref{initi}).

In order to study the time dynamics of $h (t, \vec{r})$, one can
perform a Fourier transform,
\beq
h_{{\vec k}}(t)\,=\,\frac{1}{(2 \pi)^{3/2}}\,\int
h(t,\vec{r})\,e^{ i {\vec k}\cdot \vec{r} }\,d^{3}x.
\label{Fourier}
\eeq
Now we are ready to analyze the presence (or not) of
growing modes for the particular cases. We shall present only
the final form of the equations, more details can be found
in the previous works \cite{GW-Stab} and \cite{Gasperini}. In the
last reference similar equations were obtained for the pre-Big-Bang
scenario.

To derive the wave equations we will use the
conditions (where $\,\mu\,=\,0,i=0,1,2,3$),
\beq
\partial_{i}\,h^{ij}\,=\,0
\quad \mbox{and} \quad
h_{kk}\,=\,0\,,\label{cond}
\eeq
together with the synchronous coordinate condition
$h_{\mu 0}\,=\,0$.


\subsection{Stability analysis}

In this section we will begin to analyze if there is (or not)
stability for the cosmological solutions in the theory
(\ref{4dQG}). The consideration will be based on the combination
of semi-analytical and numerical methods, where the last is mainly
used for control and illustration purposes.

The basis of the semi-analytical method is as follows. After
applying (\ref{Fourier}) we obtain a fourth-order ordinary
differential equation for the tensor part of metric perturbations.
One can easily transform it into the system of four first-order equations
and then the problem is reduced to the analysis of eigenvalues of
the corresponding characteristic equation. The details are briefly
described in Appendix B, the reader can also consult \cite{GW-Stab}.
It is easy to see that one always has to calculate the quantity
$\Delta$, given in the equation (\ref{delta}) based on the ancient
Cardano approach \cite{Cardano}. This quantity contains all relevant
information about the asymptotic behavior of the solution.

One can distinguish the following cases:
\begin{enumerate}
\item $\Delta\,<\,0$. The three roots are real and distinct. Then
we have one of the following situations:
\begin{itemize}
\item All negative roots: stable solution.
\item Some positive root: unstable and instability generally
increases with increasing number of positive roots, in a
sense one needs more severe initial conditions to avoid
instability.
\end{itemize}
\item $\Delta\,=\,0$. The roots are real,
and two or three are equal. Then
\begin{itemize}
\item All negative roots or with negative real parts: stable.
\item Some root with a positive real part: unstable
and this instability increase with increasing number
of such positive roots.
\end{itemize}
\item $\Delta\,>\,0$. One real root and two complex roots,
\begin{itemize}
\item All negative roots or with negative real parts: stable.
\item There are root with a positive real part and the solution is
unstable.
\end{itemize}
\end{enumerate}

In what follows we shall perform the analysis separately
for each case, namely for flat space-time, exponential
expansion, radiation, matter in fourth-derivative theory
and also consider the flat case for a superrenormalizable theory.
In each case we shall consider many different values of $k$ and
will try to see in which range of frequencies the growing modes
will show up.

\subsection{Flat Case}

In order to fix the method, consider first the flat case, when
$g_{\mu\nu}^{0}\,=\,(1,-\delta_{ij})$. The action of our
interest (\ref{4dQG}) can be presented as
\beq
S_{HDQG} &=& S_0+S_1+S_3\,,
\label{action}
\eeq
where
\beq
S_0\,=\,f_0\,\int d^4 x\,\sqrt{-g}\,R\,,\qquad
S_1\,=\,f_1\,\int d^4 x\,\sqrt{-g}\,C^2\,,\qquad
S_3\,=\,f_3\,\int d^4 x\,\sqrt{-g}\,R^2\,.
\eeq
The metric perturbation are defined as
\beq
g_{\mu\nu}\,=\,g^{0}_{\mu\nu} + h_{\mu\nu}
\label{Flat}
\eeq
and the synchronous and harmonic gauge fixing conditions
(\ref{cond}) are imposed. Then the second variations of
the actions yield the following results (here $h \equiv {\bar h}^\bot_{ij}$)
\beq
S_0^{(2)}
&=&
f_0\,\Big[h
\,\ddot{h}
+ \frac34\,\dot{h}\,\dot{h}
- \frac14\,h
\,\nabla^2 h\Big]
\nonumber
\\
S_1^{(2)}
&=&
f_1\,
\Big[
\frac12\,\ddot{h}^2\,
 + \frac12\, (\nabla^2 h)(\nabla^2 h)
+ \ddot{h} \, \nabla^2 h
+ 2 \dot{h} \,\nabla^2 \dot{h}\Big]
\nonumber
\\
S_3^{(2)}
&=& 0\,,
\label{varies}
\eeq
where, also $\na=\na_k,\,\,k=1,2,3$. As always, $R^2$-term does not
contribute to the tensor part of the gravitational perturbation in
flat case. Taking the sum of the three terms in (\ref{varies}), we
arrive at the equation for perturbations,
\beq
f_1\,\stackrel{....}{h}
\,- \,2 f_1 \nabla^{2} \ddot{h}
\,+\, f_1\,\nabla^4 {h}
\,+\, \frac12\, f_0\,\ddot{h}
\,-\, \frac12\, f_0 \nabla^2{h}\,=\,0\,,
\label{flat}
\eeq
that is nothing else but the equation equivalent to
(\ref{eq1-pure})\footnote {We adopt notations
\ ${h}^{l}_{k}\,{h}^{k}_{l}\,=\,h^2$, \
${h}^{l}_{k}\,\dot{h}^{k}_{l}\,=\,h\,\dot{h}$
and use $\,\Box\,{h}^{l}_{k}\,=\,\ddot{h}^{l}_{k}
- \nabla^2\,{h}^{l}_{k}$}
\beq
(f_1\,\Box^2 + f_0\,\Box)h\,=\,0\,.
\label{semBoxExtra}
\eeq

\begin{figure}
\includegraphics[height= 6.0 cm,width=10.0cm]{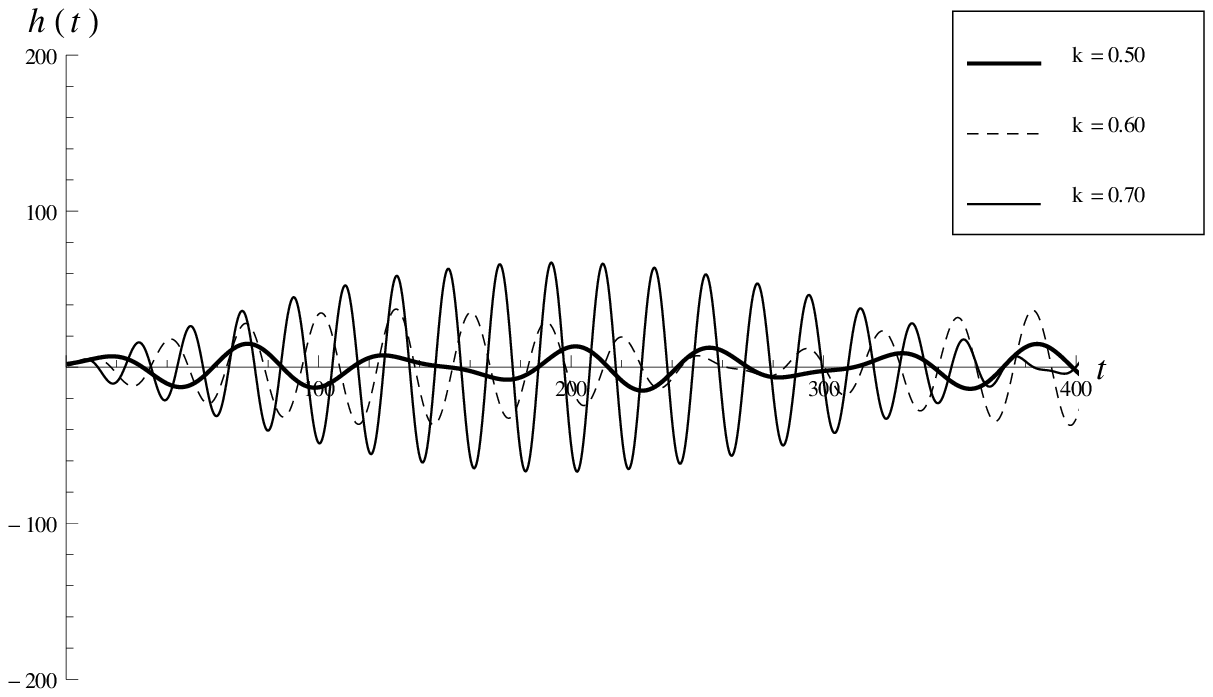}
\includegraphics[height= 6.0cm,width=10.0cm]{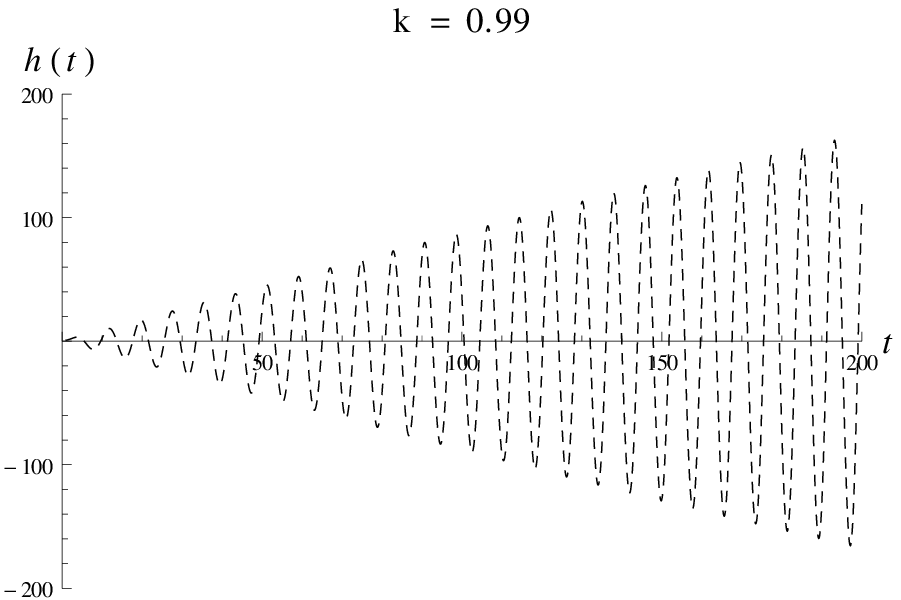}
\caption{The plots for the flat-space case. $k$ is measured in
the units of $M_P$ and the growing modes appear for $k$ close to 1.
Oscillations here means that the eigenvalues with both positive and
negative real parts have imaginary components. For smaller values
of $k$ the amplitude greatly increases, but asymptotically goes to
zero (out of the plot).}
\label{fig1}
\end{figure}

Let us present the results for the growing modes.
\begin{description}
\item[Semi-analytical analysis.]

For $a_{1}\,>\,0$ we find run-away solutions for all values of $\,k$.

For $a_{1}\,<\,0$ for $k\,<\,0.90$ we have $\,\Delta\,<\,0\,$
and all eigenvalues are real and negative. So, we have stability
in this case. For $k\,>\,0.90$ we find two positive eigenvalues.
Therefore we can observe instability, i.e., run-away solutions.

\item[Numerical analysis.] Using {\tt Mathematica}
software \cite{Wolfram}, we find that the growing modes show up
from $k \geq 0.99$. The illustrating plots for the initial period
of time are shown in Fig. \ref{fig1}.
\end{description}

One can see that for $a_1 < 0$ growing modes exist for the magnitude
of the wave vector being equal or greater than the Planck mass.
For much smaller frequencies $k$ we do not observe the effect of ghost,
probably because its mass is too large. It is important for our
general understanding that for $a_1 > 0$ there are exponentially
growing modes for all values of  $k$. In this case the massless
mode (graviton) is actually a ghost, so there is no energy gap for
generating the run-away solutions. Obviously, a huge energy gap
exists for the $a_1 < 0$ case.

Let us make one more observation concerning the marginal value
of $k$, starting from which the growing modes are observed.
According to equation (\ref{eq1-pure}), this value depends on the
ratio $M_P/\sqrt{-a_1}$. In the consideration presented above,
we have used $a_1=-1$, and consequently found that the
marginal value of $k$  is close to $M_P$.

\subsection{Cosmological solutions}

Let us now consider the dynamics of the gravitational waves
on the cosmological background. It proves useful to present
the action (\ref{4dQG}) using different notations. After
performing some integrations by parts, we arrive at
\beq
S\,=\,\int \,d^4x\sqrt{-g}\,L\,,
\label{act}
\eeq
where
\beq
L
&=&
\sum^{5}_{s\,=\,0}\,f_{s}\,L_{s}
\, = \,\Big(f_{0}\,R
+ f_{1}\,R^{\alpha \beta \mu \nu}R_{\alpha \beta \mu \nu}
+ f_{2}\,R^{\alpha \beta}R_{\alpha \beta}
+ f_{3}\,R^{2} \Big)
\label{1}
\eeq
and the coefficients $f_{0,..,3}$ are defined according to
\beq
f_{0}
&=& -\frac{{M_P}^{2}}{16\pi}\,,
\nonumber
\\
f_{1}
&=&
a_{1} + a_{2}\,,
\nonumber
\\
f_{2} &=& -2a _{1} - 4a_{2}\,,
\nonumber
\\
f_{3}\,&=&\,\frac{a _{1}}{3} + a_{2}\,.
\label{fs}
\eeq
As one should expect, the coefficient $a_2$ of the Gauss-Bonnet
topological term does not affect the equations.

Let us consider the background cosmological solution
\ $g_{\mu\nu}^{0}\,=\,\{1,-\delta_{ij}\,a^2(t)\}$. Then one can
arrive at the following expressions for the bilinear parts of
the partial Lagrangians from equation (\ref{1}),
\beq
L_{0}\,&=&\,a^{3}\,f_{0}\Big[h^{2}
\Big(\frac{3}{2}\dot{H} + 3 H^{2}\Big)
+ h\ddot{h} + 4 H h \dot{h} + \frac{3}{4} \dot{h}^{2} -
\frac{h}{4} \frac{\nabla^{2}h}{a^{2}}\Big]
+ {\cal O}(h^{3}),
\nonumber
\\
L_{1}&=&a^{3}\,f_{1}\Big[{\dot{h}}^2\big(2 H^{2}
- 2\dot{H}\big) - h\ddot{h}\big(4 H^{2}
+ 4\dot{H}\big) - h^{2}\big(3\dot{H}^{2}
+ 6\dot{H} H^{2} + 6 H^{4}\big) -
\nonumber
\\
\nonumber
&-& h\dot{h}\big(8 H \dot{H}
+ 16 H^{3}\big) + \ddot{h}^2 + 4 H \dot{h}\ddot{h} +
\Big(\frac{\nabla^{2}h}{a^{2}}\Big)^{2}
+ 2\dot{h} \frac{\nabla^{2}\dot{h}}{a^{2}} +
\\
&+& \big(H^{2} h
- 2 H \dot{h}\big)\frac{\nabla^{2}h}{a^{2}}\Big] + {\cal O}(h^{3}),
\nonumber
\\
L_{2}&=&a^{3}\,f_{2}\Big[-h\dot{h}\big(12\dot{H}H + 24 H^{3}\big)
- \frac{\dot{h}^{2}}{2}\Big(5\dot{H} + \frac{18}{4} H^{2}\Big) -
\nonumber
\\
&-& h^{2}\big(3\dot{H}^{2} + 9 \dot{H} H^{2} + 9 H^{4}\big) -
h\ddot{h}\big(4\dot{H} + 6 H^{2}\big) + \frac{\ddot{h}^{2}}{4}
+ \frac{3}{2} H \dot{h} \ddot{h}
+
\nonumber
\\
 &+&\frac{1}{4}\Big(\frac{\nabla^{2}h}{a^{2}}\Big)^{2} -
\frac{1}{2}\big(\ddot{h} + 3 H \dot{h} - \dot{H} h -
3H^{2} h\big)\frac{\nabla^{2}h}{a^{2}}\Big] + {\cal O}(h^{3}),
\nonumber
\\
L_{3}&=&- 6 a^{3}\,f_{3}\big(\dot{H} +
2 H^{2}\big)\Big[h^{2}\Big(\frac{3}{2}\dot{H} + 3 H^{2}\Big)
+ 2 h \ddot{h} +
\nonumber
\\
&+& 8 H h \dot{h} + \frac{3}{2} \dot{h}^{2}
- \frac{h}{2}\frac{\nabla^{2}h}{a^{2}}\Big] + {\cal O}(h^{3}),
\label{Ls}
\eeq
Omitting higher order terms ${\cal O}(h^{3})$ in the
expressions (\ref{Ls}) and taking variational derivative
with respect to $h_{\mu\nu}$, we arrive at the equation
for tensor mode\footnote{Which is, in fact, a part of the
more complicated equation with quantum corrections, which
was explored in \cite{GW-Stab}.},

 \beq
&& \Big( 2 f_{1} + \frac{f_{2}}{2}\Big)\, \stackrel{....}{h}
+ \big[ 3H\big(4f_1 + f_2\big) \big]\, \stackrel{...}{h}
+ \Bigl[ 3H^{2}\Big(6f_{1} + \frac{f_2}{2} - 4f_{3}\Big)
\nonumber
\\
\nonumber
&+& 6\dot{H}\big(f_1 - f_3\big)
+ \frac{1}{2} f_0 \Bigl]\,\ddot{h}
- \big(4 f_{1} + f_{2}\big)\,\frac{\nabla^2 \ddot{h}}{a^2}
\\
\nonumber
&+&
\Big[
- 21 H \dot{H} \Big(\frac{1}{2}f_2 + 2 f_3\Big)
- \ddot{H}\Big(\frac{3}{2} f_2 + 6 f_3\Big)
- 9H^{3}\big(f_2 + 4f_3\big) + \frac{3}{2} H f_0 \Big]\,\stackrel{.}{h}
\\
\nonumber
&-& H\big(4 f_{1} + f_{2}\big)\,\frac{\nabla^{2} \dot{h}}{a^{2}}
- \Big[\big(36 \dot{H} H^{2}
+ 18 \dot{H}^{2}
+ 24 H \ddot{H}
+ 4\stackrel{...}{H}\big) \big(f_{1} + f_{2} + 3 f_{3}\big)\Big]\,{h}
\nonumber
\\
&+& f_{0}\big[2 \dot{H} + 3 H^{2}\big]\,h
+ \big[H^{2} \big(4 f_{1} + 4 f_{2} + 12 f_{3}\big)
\nonumber
\\
&+& 2\dot{H}\big(f_1 + f_2 + 3 f_3\big)
- \frac{1}{2}\, f_0
\Big]\,\frac{\nabla^{2} h}{a^{2}}
+ \Big(2 f_{1}
+ \frac{1}{2} f_{2}\Big)\,\frac{\nabla^{4} h}{a^{4}}
\,=\,0\,.
\label{diff}
\eeq
This equation can be used for different cosmological
solutions. In what follows we consider three examples, namely the
exponential expansion, radiation and matter-dominated epochs.

\subsection*{Exponential expansion}

\begin{figure}
\includegraphics[height= 6.0 cm,width=10.0cm]{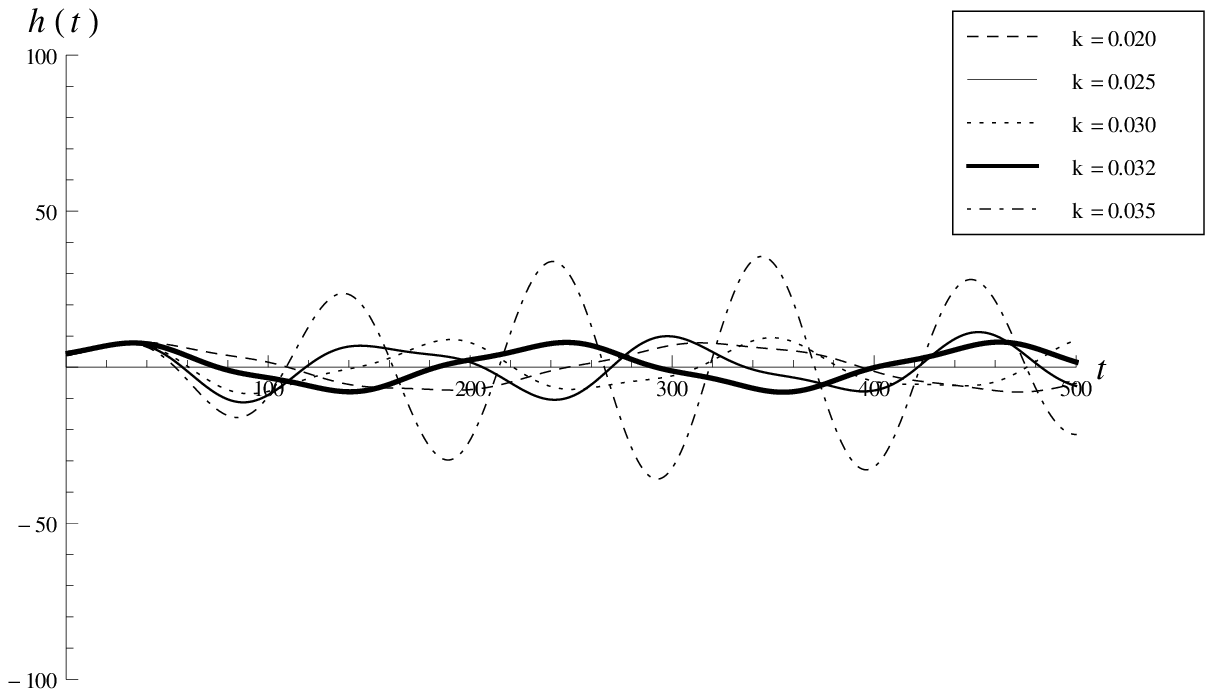}
\includegraphics[height= 6.0 cm,width=10.0cm]{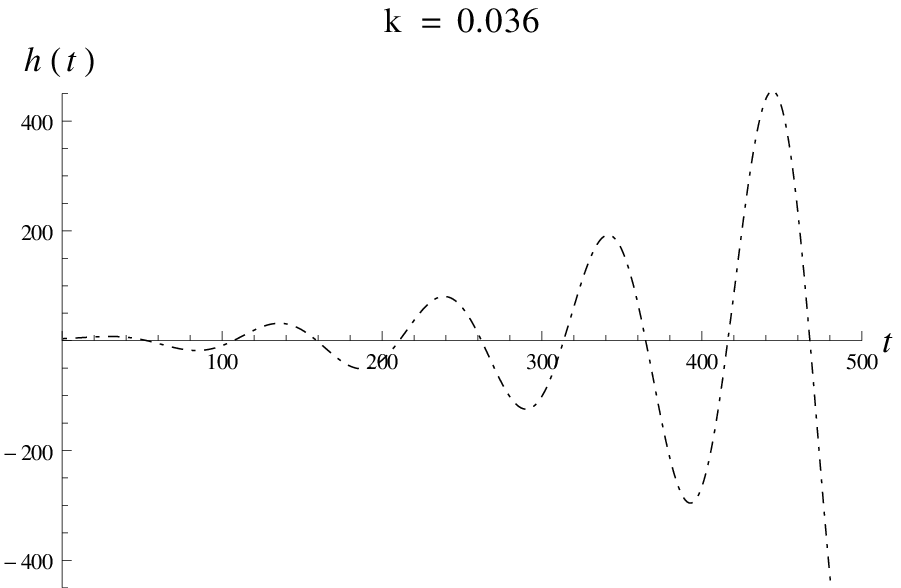}
\caption{Some plots of $\,h (t)\,$ for the exponential case,
$a(t)\,=a_{0}e^{H_{0}t}$.
The solution with growing modes appear only starting from $\,k=0.036$.}
\label{fig2}
\end{figure}

\begin{description}
\item[Semi-analytical analysis.]
For $a_{1}\,>\,0$ there are run-away solutions for all $k$ values.

In the case $a_{1}<0$, for $k<0.036$ we have $\Delta<0$
and all eigenvalues are real and negative, hence there are no
instabilities in this case. For $k\,>\,0.036$ there is one
positive eigenvalue. So, starting from this frequency one
can observe instability (i.e., run-away solutions) for the
exponential expansion of the Universe.

\item[Numerical analysis.] The result described above is
perfectly well confirmed by numerical analysis by using
{\tt Mathematica} software. For the exponential
expansion the growing modes emerge only when $k \geq 0.036$,
as it is illustrated in Fig. \ref{fig2}.
For smaller frequencies, there are no run-away solutions.
\end{description}

We can see that the effect of the non-trivial background manifest
itself mainly in the small modification of the marginal value of $k$,
after which we observe growing modes. In view of the consideration
in Sect. 3 this is an expected result, because we saw that the
weak (compared to the Planck scale) background will produce only
small corrections to $\De$ and hence to the growing modes. Let us
see whether the situation is the same for other cosmological
solutions.

\subsection*{Radiation}
\begin{figure}
\includegraphics[height= 6.0 cm,width=10.0cm]{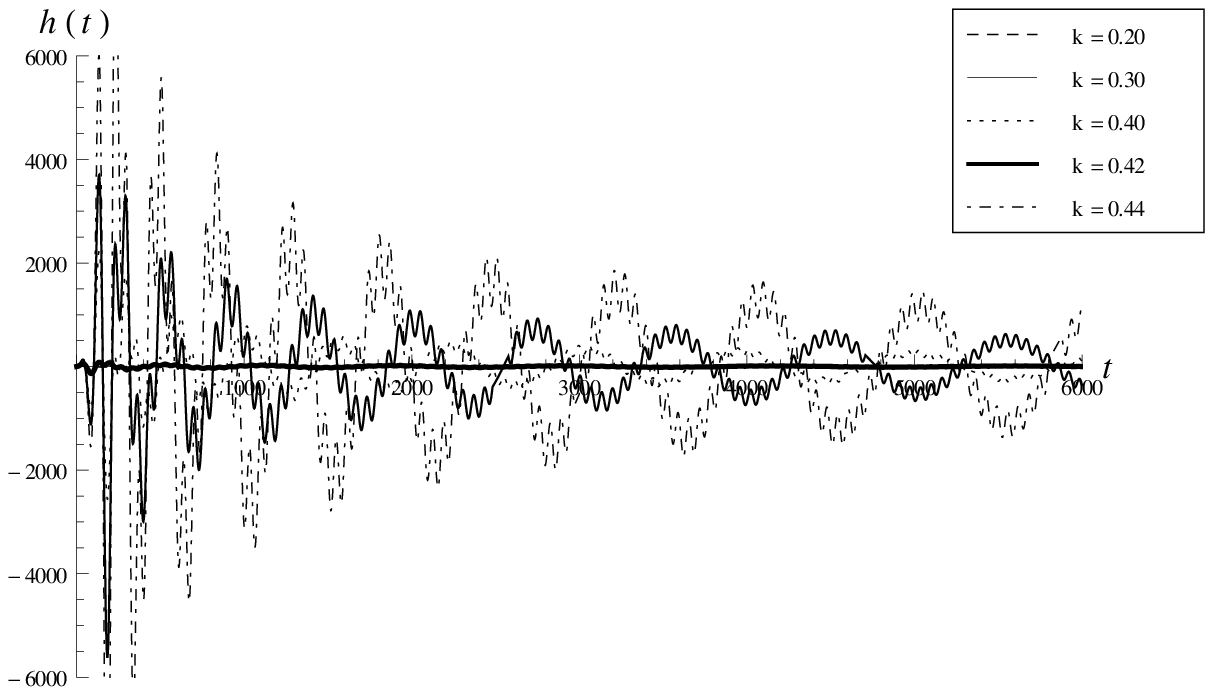}
\includegraphics[height= 6.0 cm,width=10.0cm]{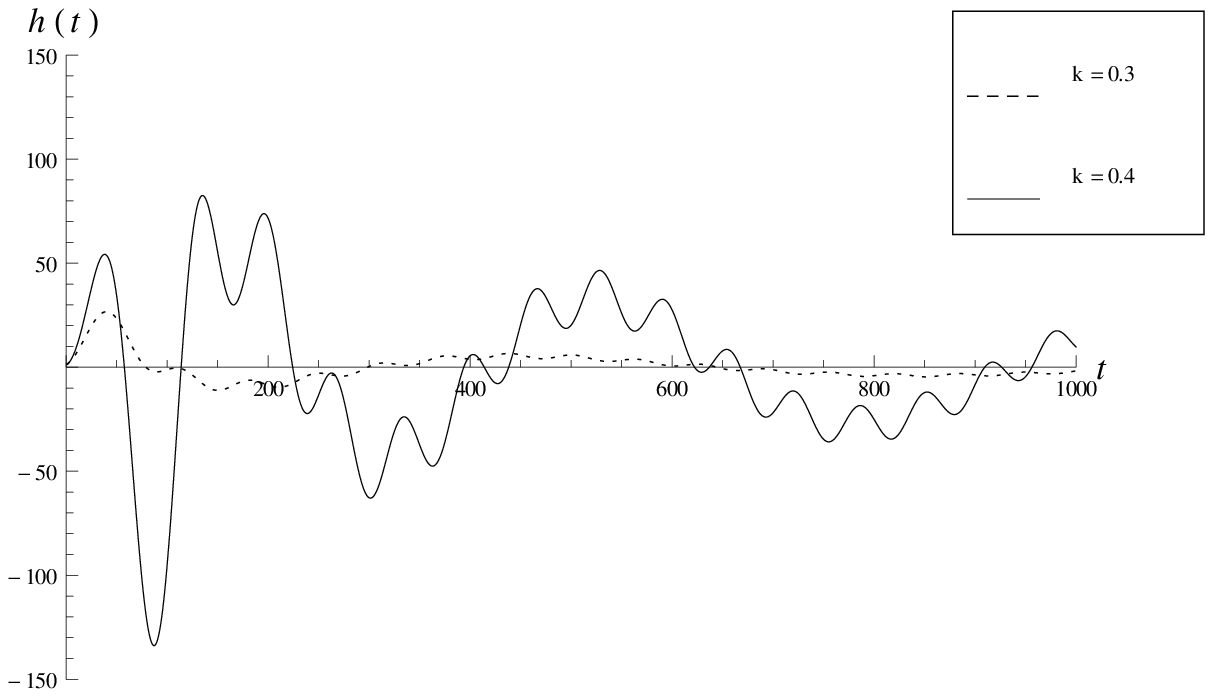}
\includegraphics[height= 6.0 cm,width=10.0cm]{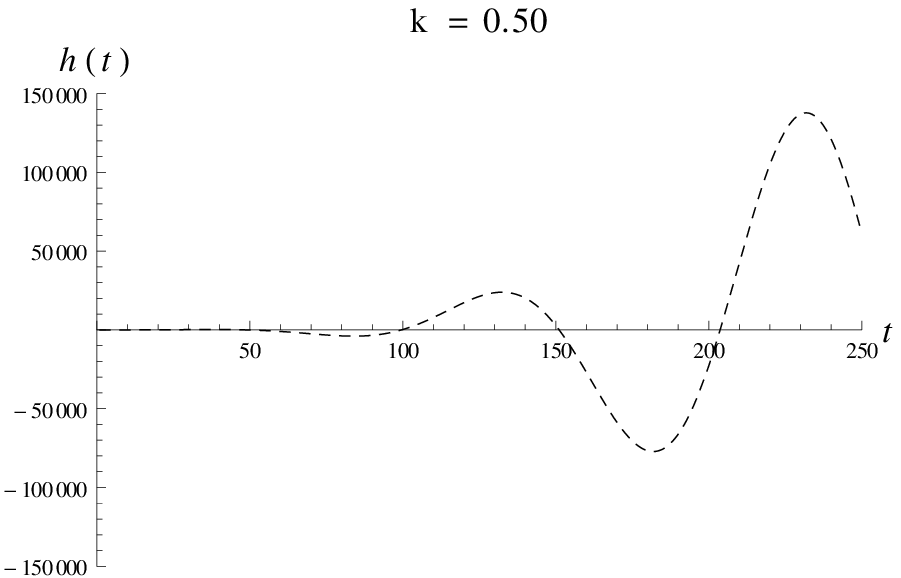}
\caption{Graph for $ h (t) $ perturbation in function of
time analyzed for the radiation, when $\,a(t)=a_{0}t^{1/2}$.
Starting from $k\,\sim\, 0.50$ the solutions become ``violent'', as o
ne can see on the last plot. However, below this value there are no
growing modes.}
\label{fig3}
\end{figure}

\begin{description}
\item[Semi-analytical analysis.]
We find run-away solutions for all $k$ values for radiation
when $a_{1}>0$, exactly as in the exponential expansion case.

If we choose $a_{1}<0$ we have $\Delta<0$ for $k<0.50$
and all eigenvalues are real and negative. Thus we have stability
for this frequency range.
But for $k>0.50$ we find extremely large values of $h(t)$
and two positive eigenvalues, so we have growing modes.

\item[Numerical analysis.] Again the results found in the semi-analytical
method agree perfectly with the analysis done by software {\tt Mathematica}.
For the case of radiation, as we can see in Fig. \ref{fig3}, we have
run-away solutions only when $ k \geq 0.44$. For smaller
frequencies we don't have this kind of solutions.
\end{description}

\subsection*{Matter}
\begin{figure}
\includegraphics[height= 6.0 cm,width=10.0cm]{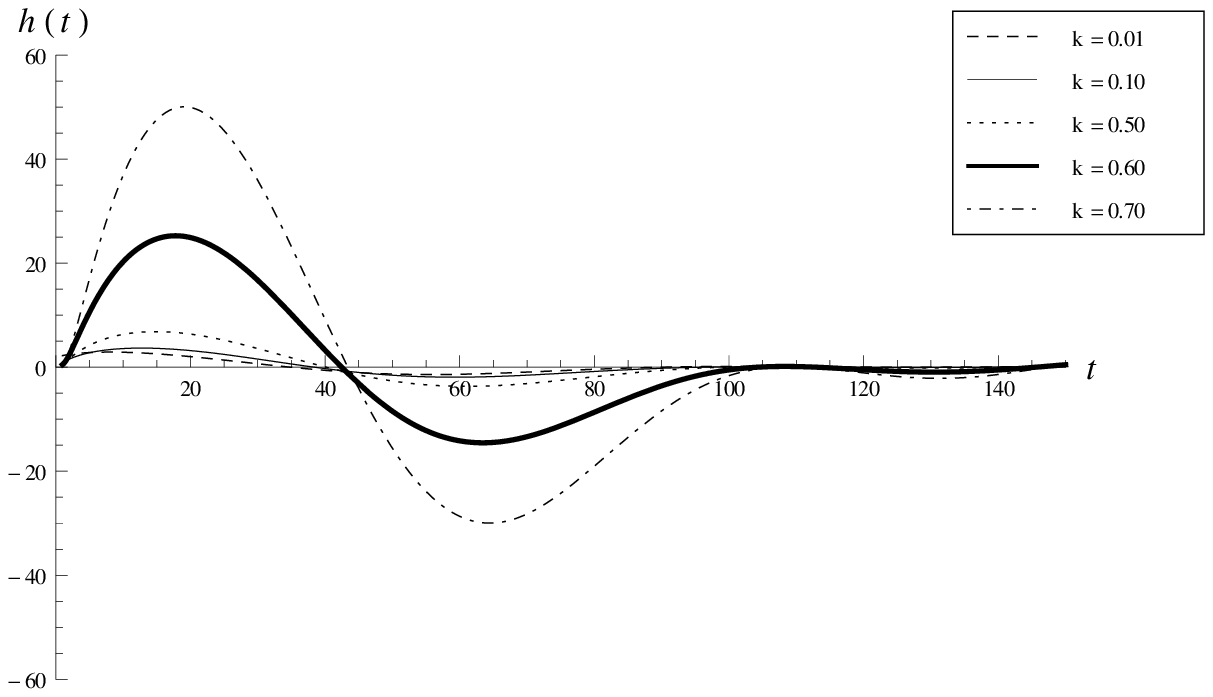}
\includegraphics[height= 6.0 cm,width=10.0cm]{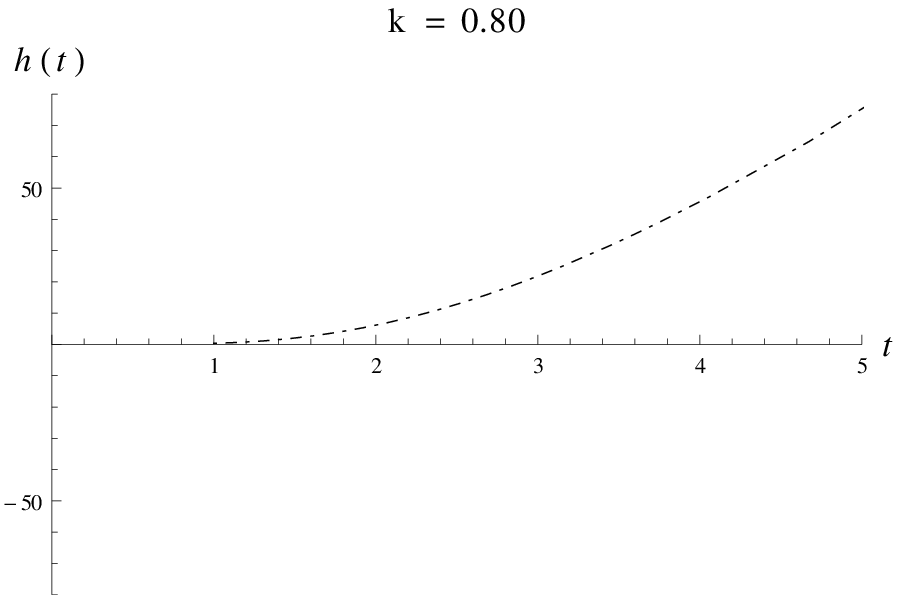}
\caption{Graph for $ h (t) $ perturbation in function of
time analyzed for $a(t)\,=a_{0}t^{3/2}$ (Matter).}
\label{fig4}
\end{figure}

\begin{description}
\item[Semi-analytical analysis.]
Once again, for $a_{1}\,>\,0$ we have run-away solutions for all
values of $k$. For $a_{1}\,<\,0$ we have $\Delta\,<\,0$ for the
$k$ values up to $k=0.80$ and all eigenvalues are real and negative,
therefore we have stability. But for $k\,>\,0.80$ we find
two positive eigenvalues, indicating to the presence of growing modes.
\item[Numerical analysis.]
Using the {\tt Mathematica}
software one can see that run-away solutions appear starting from
the frequencies $\,k \geq 1$, in a good agreement with the
semi-analytical analysis. The illustrative plots are shown in
Fig. \ref{fig4}.
\end{description}

One can see that the run-away solutions take place for smaller
values of $k$ in the case exponential expansion, then for radiation and
finally for the dust (matter). The marginal values satisfy inequalities
\beq
k\,{}_{run-away}^{Inflation}\,<\,k\,{}_{run-away}^{Radiation}
\,<\,k\,{}_{run-away}^{Matter}
\label{marg}
\eeq
However, in all cases the growing modes appear only when we have $k$
close to the Planck scale, for a negative $a_1$.


\subsection{Superrenormalizable theory}

In order to check our understaning of the relation between energy gap
for the run-away solution and the presence of massive unphysical ghost
with the Planck-order mass, let us consider the simplest possible example
of the superrenormalizable theory of gravity (\ref{higher}), by including
two next-order terms compared to the fourth-derivative theory,
\beq
S &=& S_{EH} \,+\,\int d^4x\sqrt{-g}\,\Big\{
a_1 R_{\mu\nu}^2 + a_2R^2 + \, ...
\nonumber
\\
&+& b_1 R_{\mu\nu}\Box R^{\mu\nu}
+ b_2 R\Box R
+ \, b_{4,5,..} {\cal O}(R_{...}^3) \,+
\,+\, ...\,+ \, b_{3,4,..}{\cal O}(R_{...}^3)
\Big\}\,.
\label{6ders}
\eeq
As we have already mentioned in Sect. 2, this theory has exactly the
same amount of ghosts as the fourth-derivative theory (\ref{4dQG}),
because an extra spin-two degree of freedom has positive kinetic
energy and, also, Planck-order mass. Then one should expect that
the conditions of stability in the two theories (\ref{6ders})
and  (\ref{4dQG}) should be very similar.

The consideration presented above is valid for the structure of poles,
in the spin-2 sector, according to
\beq
G_2(k) = \frac{A_{0}}{k^2} +
 \frac{A_{1}}{k^2 + m_1^2} + \frac{A_{2}}{k^2 + m_2^2}\,,
\label{poles-6ders}
\eeq
with growing real masses of poles,
\beq
0 < m_1^2 < m_2^2\,.
\label{masses-6ders}
\eeq
In this case we have $A_0>0$ and $A_2>0$, while  $A_1<0$,
according to Eq. (\ref{alter-ego}). This feature
indicates that the first massive particle, with negative sign of
$A_1$, is a ghost, while the second massive particle, with a
positive sign of $A_2$, is just a positively-defined spin-two
particle with a huge mass. From the physical side the presence
of such an extra particle can not lead to any extra instability
and this is what we intend to check here.

The first question is how to provide this structure of poles.
Let us first establish the necessary conditions for the
coefficients $a_1$ and $b_1$ in the action.
Making the expansion $g_{\mu\nu}=\eta_{\mu\nu}+h_{\mu\nu}$, we
can easily derive the bilinear terms of this action (spin-two
part only, of course) in the form
\beq
S^{(2)}_2 &=& \int d^4x\, \,h_{\mu\nu}\,
\Big(\frac{M_P^2}{64\pi}\,\Box \,-\,
\frac{a_1}{2}\,\Box^2 \,-\,\frac{b_1}{2}\,\Box^3
\Big)\,h_{\mu\nu}\,.
\label{quadra-6ders}
\eeq

For the inverse propagator we meet the expression
\beq
G_6^{-1}(k)  &=&
\frac{b_1}{2}\,k^2\,\Big(k^4 \,-\,\frac{a_1}{b_1}\,k^2
\,-\,\frac{M_P^2}{32\pi\,b_1}\Big)\,.
\label{G-16-6ders}
\eeq
The two relevant observations can be done at this moment.
First, if we want to have positive-energy graviton, the sign of $b_1$
should be positive. This is clear already from (\ref{quadra-6ders}).
Second, if we want the Planck mass to be the unique scale-defining
parameter of the theory, then the coefficient $b_1$ should be taken
as $b_1=B_1/M_P^2$, with $B_1$ being dimensionless parameter of
the order one.

With these choices, we arrive at the following representation:
\beq
G_6^{-1}(k)  &=&
\frac{b_1}{2}\,k^2\,\big(k^2 \,-\,m_1^2\Big)\,\big(k^2 \,-\,m_2^2\Big)\,,
\label{k12-6ders}
\eeq
where
\beq
m^2_{1/2} &=& M_P^2\,\Bigg[
\frac{a_1}{2B_1}
\,\mp\,\sqrt{\frac{1}{32\pi\,B_1} + \frac{a_1^2}{4B_1^2}}
\Bigg]\,.
\label{k12values-6ders}
\eeq
Obviously, one has to choose, in order to achieve the structure
of poles of (\ref{poles-6ders}), the positive sign of $a_1$,
which is opposite to the four-derivative case. Furthermore,
the inequality
\beq
a_1^2 &>& \frac{B_1}{8\pi}
\label{last}
\eeq
is requested to provide positive real poles for the propagator.
It is obvious that all these conditions can be satisfied if we
chose, for example, $a_1=B_1=1$. This will be our choice for
the given theory, with it we shall explore the time dynamics
of the gravitational perturbations in the flat case. The choice
of the flat background is natural, since it is the simplest one
and hence we avoid complications in comparison of the
stability limits for the theories (\ref{6ders}) and  (\ref{4dQG}).

The analysis of stability performs exactly like in the
fourth-derivative case, we we can directly go to the results.

\begin{figure}
\includegraphics[height= 6.0 cm,width=10.0cm]{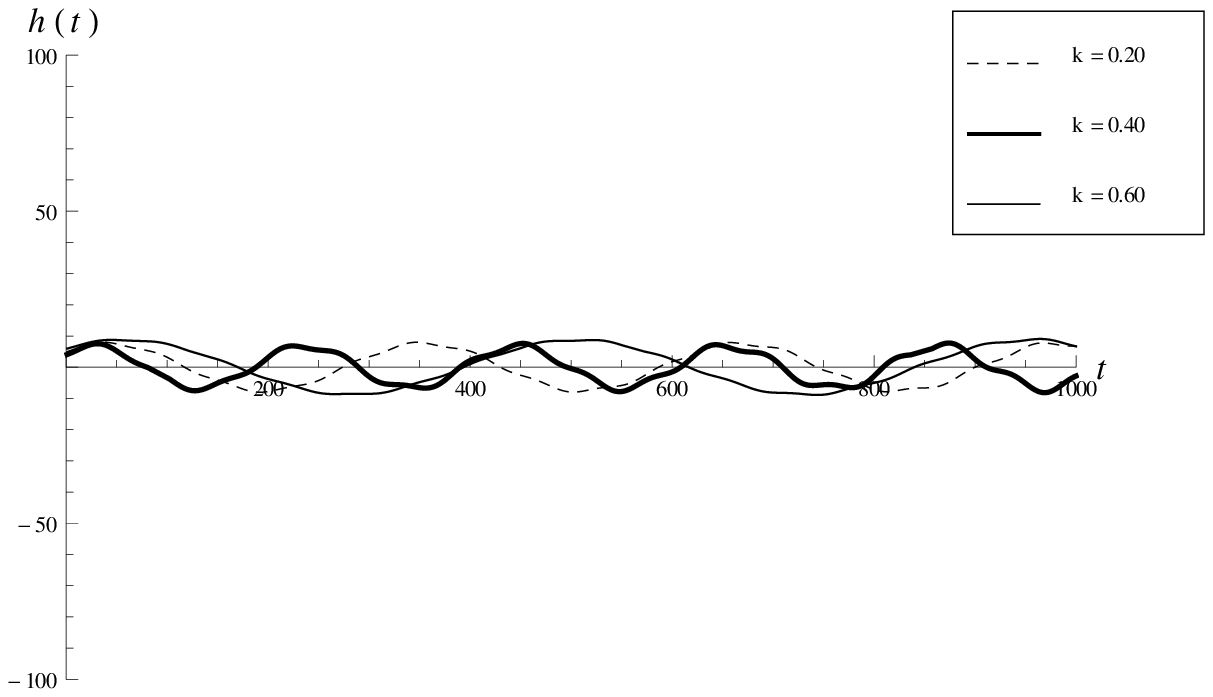}
\includegraphics[height= 6.0 cm,width=10.0cm]{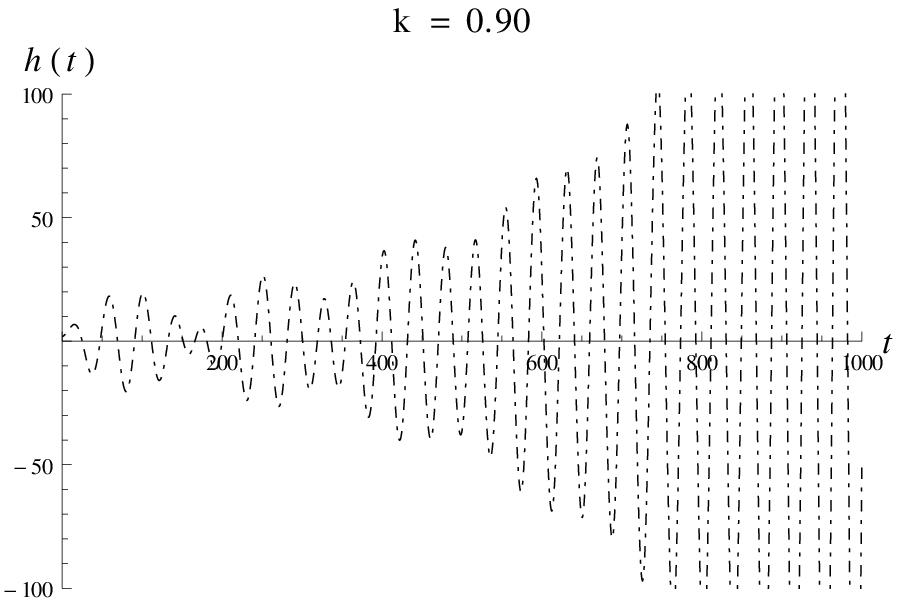}
\caption{Once again, growing modes appear only close to the Planck scale.}
\label{fig5}
\end{figure}

\begin{description}
\item[Semi-analytical analysis.]

If we choose $b_{1}\,<\,0$ we find, for $k\,<\,0.90$, that $\Delta\,<\,0$
and all eigenvalues are real and negative. So, we have stability
in this case. For $k\,>\,0.90$ we find two complex eigenvalues
with positive real parts, indicating instability.
For $a_{1}\,>\,0$ we find run-away solutions for all values of $k$.

\item[Numerical analysis.] Again, as in the cases which were
considered before, the semi-analytical method agrees with the
numerical results. In both cases there are growing modes when
$k \geq 0.90$. For the equation (\ref{k12-6ders}) the plot
is shown in Fig. \ref{fig5}. The conditions and the behavior
of perturbations look very much like in the case of the theory
(\ref{4dQG})
\end{description}

\section{Conclusions}

We considered the stability of higher derivative gravity theories
under gauge-independent part of metric perturbations. It was shown
that at least cosmological solutions are stable. Due to the
similarity with the general situation, as it is described in Sect. 3,
it might happen that this is true for any classical solution. The
perturbations which we have dealt with
were taken at the linear level, but over the non-trivial metric
background, so according to the known theorems \cite{Math} the
linear stability should be a sufficient condition of the stability
even beyond the linear approximation, if the amplitude of initial
perturbations is sufficiently small.

One can make two natural questions concerning this situation.

First, as we have already mentioned in the Introduction, any kind
of classical solution is obviously not protected by energy
conservation from the process in which one massive ghost and large
amount of gravitons are created at the same time. So, the first
question is how one can reconcile this with the stability
properties? Let us confess that we have no definite answer to
this question. At the same time physical intuition tell us
that the situation when we need to accumulate a Planck-order
energy density of gravitons in the vicinity of certain space-time
point, where the ghost should be created, means that we go to
the physics at Planck anergy scale. As far as we intend to have
a consistent QG theory at the energy scale a few orders beyond
$M_P$, there is a hope to achieve a consistent solution to this
discrepancy. For example, in the recent papers \cite{DG} one can
find a discussion of the possible limits on the occupation number
of gravitons in a gravitational field. It might happen that such
limits can be very useful for understanding the situation with
creation of ghosts from vacuum in higher derivative QG.
Furthermore, we can not rule out that the solution of the
problem can be achieved even for the Planck scale of energy,
if we better understand the physical principles behind such
limits.

Second, are the cosmological solutions sufficiently general
to draw general conclusions? In our opinion the answer is
negative. We mainly dealt with these solutions because they
are the simplest ones and the technique of corresponding
perturbations is better developed. At the same time, it would
be very interesting to explore, using effective framework,
the stability of the static black hole metric, where we have
contradicting results \cite{Whitt} and \cite{Myung}. It would
be very important to have certain results on the stability of
this and other relevant solutions, e.g., for the Kerr metric.

Finally, let us note that one single definite example of unstable
physically relevant solution in the theory with higher derivatives
would mean that the situation with the (in)stability of vacuum in
this theory becomes definitely negative. In view of the great
relevance of higher derivatives, especially for quantization of
matter fields on curved background, this would mean the necessity
of some dramatic changes in our understanding, starting from the
semiclassical approach to gravity. However, after considerations
presented in this work, we have an expectation that the situation
with higher derivatives in a theory based on a unique Planck scale
can be resolved.

\appendix

\section{Second-order expansions}

The expansion for the metric has the form
\beq
g_{\al\be}(y)
&=& {\stackrel{\mbox{\tiny o}}{g}}_{\al\be}
\,-\,\frac13\,{\stackrel{\mbox{\tiny o}}{R}}_{\mu\al\nu\be}
\,y^\mu\,y^\nu
\,-\,\frac{1}{3!}\,
{\stackrel{\mbox{\tiny o}}{R}}_{\mu\al\nu\be\,;\,\si}
\,y^\mu\,y^\nu\,y^\si
\nonumber
\\
&+&
\frac{1}{5!}\,\Big(
\frac{16}{3}\,
{\stackrel{\mbox{\tiny o}}{R}}\,^\la_{\,\,\cdot\,\mu\al\nu}
\,{\stackrel{\mbox{\tiny o}}{R}}_{\la\rho\be\si}
\,-6\,{\stackrel{\mbox{\tiny o}}{R}}_{\al\mu\be\nu\,;\,\rho\si}
\Big)
\,y^\mu\,y^\nu\,y^\rho\,y^\si
+\,...\,.
\label{metric2-A}
\eeq
One can always choose the metric in the expansion point to be
Minkowski one, \ ${\stackrel{\mbox{\tiny o}}{g}}_{\al\be} =
\eta_{\al\be}$. For the Christoffel symbol one has
\beq
\nonumber
\Gamma^{\lambda}_{\alpha\beta}
&\approx&
\frac{2}{3}\,
{\stackrel{\mbox{\tiny o}}{R}}\,^{\la}_{\,\,\,(\al\be)\nu} \,y^{\nu}
\\
&+& \frac18
\Big({\stackrel{\mbox{\tiny o}}{R}}\,^{\la}_{\,\,\,\mu\nu\be;\,\al}
+ {\stackrel{\mbox{\tiny o}}{R}}\,^{\la}_{\,\,\,\al\nu\be;\,\mu}
+ 2{\stackrel{\mbox{\tiny o}}{R}}\,^{\la}_{\,\,\,\be\mu\al;\,\nu}\Big)
\,y^\mu\,y^{\nu}\,.
\end{eqnarray}

Let us present the results of the expansions for \ $\Box\,h^{\al\be}$.
We will label by $A^{(0)}$ the order of expansion in \ $y^\mu$ for the
quantity $A$, such that
\beq
\Box\,h^{\alpha\beta}\,=\,(\Box\,h^{\alpha\beta})^{(0)} +
(\Box\,h^{\alpha\beta})^{(1)} + (\Box\,h^{\alpha\beta})^{(2)}\,+\,\,...
\eeq
The direct calculation yields the following results in zero and
first order in the deviation $y^\mu$:
\beq
(\Box\,h^{\alpha\beta})^{(0)}
&=&
\eta^{\mu\nu}\Big(\partial_\mu \partial_\nu\, h^{\alpha\beta}
- \frac13\,{\stackrel{\mbox{\tiny o}}{R}}\,^{\al}_{\,\,\,\nu\la\mu}
\,h^{\lambda\beta}
- \frac13\,{\stackrel{\mbox{\tiny o}}{R}}
\,^{\be}_{\,\,\,\nu\la\mu}\,h^{\al\la}\Big)\,,
\label{h0-A}
\eeq
and
\beq
(\Box\,h^{\alpha\beta})^{(1)}
&=&
\eta^{\mu\nu}\Bigg[-\frac{2}{3}\Big({\stackrel{\mbox{\tiny o}}{R}}
\,^{\al}_{\,\,\nu\la\chi} +
{\stackrel{\mbox{\tiny o}}{R}}
\,^{\al}_{\,\,\la\nu\chi}\Big)\,\pa_\mu\,h^{\la\beta}
- \frac23\,\Big({\stackrel{\mbox{\tiny o}}{R}}\,^{\be}_{\,\,\nu\la\chi}
+ {\stackrel{\mbox{\tiny o}}{R}}
\,^{\be}_{\la\nu\chi}\Big)\,\pa_\mu\,h^{\al\la}
\nonumber
\\
&+& \frac{1}{4}\,\Big({\stackrel{\mbox{\tiny o}}{R}}
\,^{\al}_{\,\,\chi\mu\la ; \nu}
+ {\stackrel{\mbox{\tiny o}}{R}}\,^{\al}_{\nu\mu\la ; \chi}
+ 2 {\stackrel{\mbox{\tiny o}}{R}}
\,^{\al}_{\,\,\la\chi\nu ;\mu}\Big)\,h^{\la\be}
+ \frac{2}{3}\,{\stackrel{\mbox{\tiny o}}{R}}
\,^{\la}_{\mu\nu\chi}\,\pa_{\la}\,h^{\al\be}
\nonumber
\\
&+&
\frac{1}{4}\Big({\stackrel{\mbox{\tiny o}}{R}}
\,^{\be}_{\,\,\chi\mu\la ; \nu}
+ {\stackrel{\mbox{\tiny o}}{R}}
\,^\be_{\,\,\nu\mu\la ; \chi}
+ 2 {\stackrel{\mbox{\tiny o}}{R}}
\,^\be_{\,\,\la\chi\nu ; \mu}\Big)\,h^{\al\la} \Bigg]\,y^{\chi}\,,
\label{h1-A}
\eeq
where semicolon indicates covariant derivative taken at the point
$P$. \ Furthermore, in the second order in $y^\mu$ we meet
\beq
(\Box\,h^{\al\be})^{(2)}
&=&
\frac{1}{3}\,
{\stackrel{\mbox{\tiny o}}{R}}\,^{\mu\,\,\,\nu}_{\,\,\,\chi\,\,\,\om}
\Bigg\{\partial_\mu \partial_\nu h^{\alpha\beta} - \frac{1}{3}\,
\big({\stackrel{\mbox{\tiny o}}{R}}
\,^{\al}_{\,\,\nu\la\mu}
- {\stackrel{\mbox{\tiny o}}{R}}\,^{\al}_{\,\,\la\nu\mu}\big)\,h^{\la\be}
- \frac{1}{3}\,
\big({\stackrel{\mbox{\tiny o}}{R}}^{\be}_{\,\,\nu\lambda\mu}
- {\stackrel{\mbox{\tiny o}}{R}}
\,^{\be}_{\,\,\la\nu\mu}\big)
\,h^{\al\la}\Bigg\}\,y^{\chi}\,y^{\om}
\nonumber
\\
&+&\eta^{\mu\nu}\Bigg\{\frac{1}{8}\,
\Big( {\stackrel{\mbox{\tiny o}}{R}} \,^\al_{\,\,\chi\om\la ; \nu}
+ {\stackrel{\mbox{\tiny o}}{R}}
\,^\al_{\,\,\nu\om\la ; \chi}
+ 2\,{\stackrel{\mbox{\tiny o}}{R}}
\,^{\al}_{\,\,\la\chi\nu ; \om}\Big)\pa_\mu\,h^{\lambda\beta}
\nonumber
\\
&+& \frac{1}{8}\Big(
{\stackrel{\mbox{\tiny o}}{R}}\,^{\be}_{\,\,\chi\om\la ; \nu}
+  {\stackrel{\mbox{\tiny o}}{R}}
\,^{\be}_{\,\,\nu\om\la ; \chi}
+ 2\, {\stackrel{\mbox{\tiny o}}{R}}
\,^{\be}_{\,\,\la\chi\nu ; \om}\Big)\partial_\mu\,h^{\al\la}
\nonumber
\\
&-& \frac{1}{8}\Big( {\stackrel{\mbox{\tiny o}}{R}}
\,^{\la}_{\,\,\chi\om\la ; \mu}
+ {\stackrel{\mbox{\tiny o}}{R}}\,^{\la}_{\,\,\mu\om\nu ; \chi}
+ 2\,
{\stackrel{\mbox{\tiny o}}{R}}\,^{\la}_{\,\,\nu\chi\mu ; \om}\Big)
\pa_\lambda\,h^{\al\be}
\nonumber
\\
&-& \frac{2}{9}\,
{\stackrel{\mbox{\tiny o}}{R}}
\,^{\la}_{\,\,\mu\nu\chi}\Big[(
{\stackrel{\mbox{\tiny o}}{R}}
\,^{\al}_{\,\,\la\tau\om}
+ {\stackrel{\mbox{\tiny o}}{R}}
\,^{\al}_{\,\,\tau\la\om})\,h^{\tau\beta}
+ ({\stackrel{\mbox{\tiny o}}{R}}
\,^{\be}_{\,\,\la\tau\om}
+ {\stackrel{\mbox{\tiny o}}{R}}
\,^{\be}_{\,\,\tau\la\om})\,h^{\alpha\tau}\Big]
\nonumber
\\
&+& \frac{1}{8}\,
\Big({\stackrel{\mbox{\tiny o}}{R}}\,^\al_{\,\,\chi\om\la ; \mu}
+ {\stackrel{\mbox{\tiny o}}{R}}\,^{\al}_{\,\,\mu\om\la ; \chi}
+ 2\,{\stackrel{\mbox{\tiny o}}{R}}\,^\al_{\,\,\la\chi\mu ; \om}
\Big) \,\pa_\nu\,h^{\la\be}
\nonumber
\\
&+& \frac19\,\big({\stackrel{\mbox{\tiny o}}{R}}
\,^{\al}_{\,\,\mu\la\chi}
+ {\stackrel{\mbox{\tiny o}}{R}}
\,^{\al}_{\la\mu\chi}\big)
\Big[\big({\stackrel{\mbox{\tiny o}}{R}}\,^{\la}_{\nu\tau\om}
+ {\stackrel{\mbox{\tiny o}}{R}}\,^{\la}_{\,\,\tau\nu\om}\big)
h^{\tau\beta}
+ \big({\stackrel{\mbox{\tiny o}}{R}}\,^{\be}_{\,\,\nu\tau\om}
+ {\stackrel{\mbox{\tiny o}}{R}}
\,^{\be}_{\,\,\tau\nu\om}\big)h^{\la\tau}\Big]
\nonumber
\\
&+& \frac{1}{9}\big(
{\stackrel{\mbox{\tiny o}}{R}}\,^{\be}_{\,\,\mu\la\chi}
+ {\stackrel{\mbox{\tiny o}}{R}}
\,^{\be}_{\,\,\la\mu\chi}\big)
\Big[\big({\stackrel{\mbox{\tiny o}}{R}}\,^{\la}_{\,\,\nu\tau\om}
+ {\stackrel{\mbox{\tiny o}}{R}}\,^{\la}_{\,\,\tau\nu\om}\big)
h^{\alpha\tau}
+ \big({\stackrel{\mbox{\tiny o}}{R}}\,^{\al}_{\,\,\nu\tau\om}
+ {\stackrel{\mbox{\tiny o}}{R}}
\,^{\al}_{\,\,\tau\nu\om}\big)
h^{\la\tau}\Big]
\nonumber
\\
&+& \frac{1}{8}
\Big(
{\stackrel{\mbox{\tiny o}}{R}}\,^{\be}_{\,\,\chi\om\la ; \mu}
+ {\stackrel{\mbox{\tiny o}}{R}}\,^{\be}_{\,\,\mu\om\la ; \chi}
+ 2\,{\stackrel{\mbox{\tiny o}}{R}}
\,^{\be}_{\,\,\la\chi\mu ; \om}\Big)
\pa_\nu\,h^{\al\la}\Bigg\}\,y^{\chi}\,y^{\om}
\label{h2-A}
\eeq

Alltogether, we find
\beq
\Box\,h^{\al\beta}
&=&
(\Box\,h^{\al\be})^{(0)}
\,+\,(\Box\,h^{\al\be})^{(1)}
\,+\,(\Box\,h^{\al\be})^{(2)}
\,+\,...\,,
\label{uravn-A}
\eeq
where the dots indicate to the terms of higher orders in $y^\mu$ and
terms of higher orders in curvature tensor and its covariant derivatives
at the point $P$.

\section{Backgrounds of our semi-analytical method}

We found the following type of fourth-order differential equation
for tensor perturbations,
\beq
      b_{4}\,\stackrel{....}{h}
\,+\, b_{3}\, \stackrel{...}{h}
\,+\, b_{2}\, \stackrel{..}{h}
\,+\, b_{1}\, \stackrel{.}{h}
\,+\, b_{0}\, \stackrel{}{h} \,=\,0\,,
\label{mathematica}
\eeq
where $b_0$, $b_1$, $b_2$, $b_3$ and $b_4$ are the coefficients
of this equation. One can reduce this fourth-order equation to
a system of four first-order equations.
Changing the variables, we have,
\beq
h_{0}\,=\,h\,,\quad
h_{1}\,=\,\dot{h}_{0}\,=\,\dot{h}
\,,\quad
h_{2}\,=\,\dot{h}_{2}\,=\,\ddot{h}
\,,\quad
h_{3}\,=\,\dot{h}_{2}\,=\,\stackrel{...}{h}\,.
\label{h1234}
\eeq
Now we can rewrite as
\beq
\nonumber
\dot{h}_{3}
&=& -\, \frac{1}{b_4}\,\Big(b_{3}\,{h}_{3} +
 b_{2}\,{h}_{2} + b_{1}\,{h}_{1} + b_{0}\,{h}_{0}\Big)\,,
\\
\nonumber
\dot{h}_{2}\,&=&\,h_{3}\,,
\\
\nonumber
\dot{h}_{1}\,&=&\,h_{2}\,,
\\
\nonumber
\dot{h}_{0}\,&=&\,h_{1}\,.
\eeq
Rewriting the differential equation, we arrive at
\beq
\nonumber
\dot{h}_{3}
&=& -\, \frac{1}{b_4}\,\Big(b_{3}\,{h}_{3} +
 b_{2}\,{h}_{2} + b_{1}\,{h}_{1} + b_{0}\,{h}_{0}\Big)\,,
\\
\nonumber
\dot{h}_{2}\,&=&\,h_{3}\,,
\\
\nonumber
\dot{h}_{1}\,&=&\,h_{2}\,,
\\
\nonumber
\dot{h}_{0}\,&=&\,h_{1}\,.
\eeq
We can rewrite this linear system of four equations
in a matrix form and easily compute the eigenvalues and
eigenvectors.
Thus, we can write in simplified form,
\beq
\dot{h}_{k}\,=\,A^l_k\,h_l\,,
\eeq
where $k\,=\,0,1,2,3$ and the matrix $A=A^l_k$ has the form
$$
A\,=\,\left(
\begin{array}{cccc}
0 & 1 & 0 & 0 \\
0 & 0 & 1 & 0 \\
0 & 0 & 0 & 1 \\
d_{0} & d_{1} & d_{2} & d_{3} \\
\end{array}
\right)\,,
$$
we called \ $d_{k}\,=\,-b_{k}/b_{4}$\,.
We need find the eigenvalues of $A$ and for this
end we consider
\beq
det\,\left(
\begin{array}{cccc}
-\lambda & 1 & 0 & 0 \\
0 & -\lambda & 1 & 0 \\
0 & 0 & -\lambda & 1 \\
d_{0} & d_{1} & d_{2} & (d_{3} -\lambda)\\
\end{array}
\right)\,=\,0\,.
\eeq
The algebraic equation is
\beq
\lambda^{4} - d_{3}\,\lambda^{3} - d_{2}\,\lambda^{2} -
d_{1}\,\lambda^{1} -d_{0}\,=\,0\,.
\label{polynomial}
\eeq
After some algebraic operation due to Cardano \cite{Cardano}
one can reduce this fourth-order equation to the second-order one,
\beq
z^{2} + \xi_{1}\,z + \xi_{2}\,=\,0\,,
\eeq
where the most important quantity is given by
\beq
\Delta\,=\,\xi_{1} + \,\frac{4}{27}\,\xi_{2}^{3}\,=\,
4\Biggl[\Bigl(\frac{\xi_{1}}{2}\Bigl)^{2} +
\Bigl(\frac{\xi_{2}}{3}\Bigl)^{3}\Biggl]\label{delta}\,.
\eeq
The value of $ \Delta $  will tell us the nature of these roots,
as explained in the text. To find equation (\ref{delta}) we use that

\beq
\xi_{1} &=& \frac{-\alpha}{3} + \beta
\quad
\mbox{and}
\quad
\xi_{2}\,=\,\Biggl(\frac{2\alpha^{3}}{27} +
\frac{3\gamma - \beta\,\gamma}{3}\Biggl)\,,
\nonumber
\\
\al &=& \frac{5}{2}\,p
\,;\quad
\gamma\,=\,\frac{1}{8}\Big(q^{2} - 4 p^{2}
+ 4\,p\,r\Big)
\quad \mbox{and} \quad
\beta\,=\,2 p^{2} - r\,,
\label{condi}
\\
p &=& - \frac{39}{8}\,d_{3}^{2} + d_{2}\,;
\quad
q\,=\,\frac{d_{3}^{2}}{8} - \frac{d_{3}d_{2}}{2} + d_{1}
\quad
\mbox{and}
\quad
r\,=\,-\frac{3d_{3}^{4}}{256} +
\frac{d_{2}d_{3}^{2}}{16} - \frac{d_{2}d_{1}}{4} + d_{0}\,.
\nonumber
\eeq
where $b_{k}/b_{4}\,=\,-d_{k}$, and $b_{k}$ are the coefficients
of the fourth-order differential equation.

\begin{acknowledgments}
One of the authors (I.Sh.) is grateful to R. Balbinot,
P. Creminelli for useful (including critical) discussions.
Authors acknowledge the contribution of Grigori Chapiro
who gave us useful hints concerning semi-analytical methods
for exploring stability. We are grateful to Sebastiao Mauro
for the discussion of the black hole case. Furthermore,
authors are grateful to CAPES, CNPq and FAPEMIG for partial support.
I.Sh. was also supported by ICTP Associate Membership program.
\end{acknowledgments}


\section*{\Large Erratum in `Do we have unitary and (super)renormalizable
Quantum Gravity below Planck scale?'}

\vspace{1cm}

\begin{center}
 Filipe de O. Salles\footnote[1]{Universidade Federal de Juiz de Fora} and Ilya L. Shapiro\footnote[2]{Universidade Federal de
 Juiz de Fora and also at Tomsk State
Pedagogical University, Tomsk, Russia.}
\end{center}

\vspace{1cm}

\begin{center}
{\bf Abstract:} We correct the formulation of physical interpretation regarding
the stability of metric perturbations in higher derivative theories
of gravity.
\end{center}

\vspace{1,5cm}

In this work (Phys.\ Rev.\ D {\bf 89}, 084054 (2014)
we have considered the stability of the cosmological
solutions in the higher derivative theory with the action
\beq
S_{4dQG}\,=\,
\int d^4x\sqrt{-g}\,\left\{
-\,\frac{\ka}{16\pi}\,R\,
\,+\,a_1C^2 + a_2E + a_3{\Box}R + a_4R^2\right\}\,.
\eeq
In the case $a_1<0$ and $\ka=M_P^2>0$ we have correctly found
that the grows of the metric perturbations starts only for the
initial frequencies of the Planck order of magnitude. Also, we
have correctly notes that for $a_1>0$ and $\ka=M_P^2>0$ there
is no Planck threshold, and the overproduction of gravitons
takes place for any initial frequency. However, the correct
interpretation of this instability is the emergence of tachyon
modes for the negative-square mass of the field. The case which
we described in the paper corresponds to the situation
$a_1>0$ and $\ka=-M_P^2<0$, when massless graviton becomes
ghost and massive graviton is a normal field. Now, we have
additionally checked, using the method described in I, that
our conclusions remain completely correct, also, in this case.

\vspace{2cm}

{\bf Acknowledgments:} Authors are thankful to Prof. M. Maggiore for
clarifying discussion of the issue. We are grateful to  
FAPEMIG and (I.Sh.) to CNPq and ICTP for partial support.

\end{document}